\documentclass[%
twocolumn,
superscriptaddress,
onlinecite,
 amsmath,amssymb,
 aps,
prb,
]{revtex4-1}

\usepackage{graphicx}
\usepackage{dcolumn}
\usepackage{comment}
\usepackage[dvipsnames]{xcolor}
\usepackage{hyperref}
\usepackage{bm}
\usepackage{amsmath}
\usepackage[normalem]{ulem}

\hypersetup{
	colorlinks = true,
	pdftitle = {},
	pdfauthor = {},
	pdfkeywords = {},
	linkcolor = blue,
	citecolor = blue,
	filecolor = black,
	urlcolor = blue
}

\usepackage{color}

\newcommand{\dd}{\mathrm{d}}
\newcommand{\ee}{\mathrm{e}}

\begin{document}

\title{Magnetoelectricity of Topological Solitons in 2D Magnets}

\author{Alexander Edstr\"om}
\affiliation{Department of Applied Physics, School of Engineering Sciences,KTH Royal Institute of Technology, AlbaNova University Center, 10691 Stockholm, Sweden}

\author{Paolo Barone}
\affiliation{Consiglio Nazionale delle Ricerche CNR-SPIN,  Area della Ricerca di Tor Vergata,Via del Fosso del Cavaliere 100, I-00133 Rome, Italy}

\author{Silvia Picozzi}
\affiliation{Department of Materials Science, University of Milan-Bicocca, 20125 Milan, Italy}
\affiliation{Consiglio Nazionale delle Ricerche CNR-SPIN, c/o Universit\`{a} degli Studi 'G. D'Annunzio', 66100, Chieti, Italy}

\author{Massimiliano Stengel}
\affiliation{Institut  de  Ci\`encia  de  Materials  de  Barcelona  (ICMAB-CSIC),  Campus  UAB,  08193  Bellaterra, Spain}
\affiliation{ICREA  -  Instituci\`{o}  Catalana  de  Recerca  i  Estudis  Avan\c{c}ats,  08010  Barcelona,  Spain}

\begin{abstract}
We develop a multiscale approach to magnetoelectric effects, bridging atomistic and continuum models, with all parameters determined from \emph{ab initio} electronic structure calculations. We show that the parameters of the model are equivalent to the electric field-induced Dzyaloshinski-Moriya interactions. After careful validation, we apply the models to study electric polarization and dipole moments carried by spin spirals and topological solitons, in the form of magnetic domain walls and Skyrmions, in the prototypical 2D magnet CrI$_3$. We show that the reduced symmetry of the material leads to additional magnetoelectric coupling terms, dominating over those expected in high symmetry (cubic) materials. An interesting consequence is that Skyrmions carry an out-of-plane electric dipole moment, while that of anti-Skyrmions is an order of magnitude larger and in-plane. Finally, we discuss the possibility to stabilize non-collinear spin states using electric fields. 
\end{abstract}

\maketitle

\section{Introduction}
The linear magnetoelectric effect, manifesting itself as a bilinear coupling between magnetization and electric polarization, was first predicted~\cite{Dzyaloshinskii_magnetoel} and observed~\cite{Astrov_Cr2O3} in Cr$_2$O$_3$. Later, a range of magnetoelectric effects describing the coupling between magnetic and electric order parameters have been investigated, including inhomogeneous ones~\cite{inhom_magnetoel}, where a spatially varying magnetic order induces an electric polarization. This effect has been widely studied in the field of multiferroics~\cite{doi:10.1126/science.1113357,Spaldin2019}, where an electric polarization is induced by spin spiral states in so called type II multiferroics~\cite{khomskii,Tokura_spiralmulti}. Conversely, applied electric fields, and the resulting electric polarization, induce Dzyaloshinskii-Moriya interactions~\cite{PhysRevB.73.094434} which stabilize non-collinear spin textures in otherwise collinear ferromagnets. 
Due to the same mechanisms, other inversion symmetry breaking magnetic structures can also generate an electric polarization coupling them to electric fields. For example, topological solitions, such as magnetic domain walls (DWs)~\cite{Hubert_DW}, which separate regions with different orientations of magnetization, can carry an electric dipole moment~\cite{khomskii}. Magnetic DWs are technologically interesting as functional units, e.g. for racetrack memories~\cite{racetrack} and the possibility of controlling them by electrical means is attractive in this context. 
Also Skyrmions~\cite{Bogdanov2020,Tokura2021}, another form of topologically protected magnetic soliton with widespread interest, can carry a local polarization and net electric dipole moment. While many of the early Skyrmion materials studied were metallic, recent examples include insulating ones where an electric polarization has been  observed.~\cite{Seki_skyrmions,doi:10.1126/sciadv.1500916} This finding enabled electric field control of Skyrmions~\cite{Okamura2016}, inspiring further efforts to stabilize Skyrmions with electric fields~\cite{Hsu2017,Behera_CrI3_Efield,PhysRevB.104.L060409,PhysRevB.108.134430}, even in antiferromagnets (AFMs)\cite{Chaudron2024}. Topological magnetic solitons, such as DWs or Skyrmions, are spatially localized but extend over length scales of at least several nanometers or more, leaving them beyond reach of direct first principles calculations, and hampering a thorough theoretical understanding and efficient design of functional materials. A few earlier works made theoretical estimates of the polarization of atomically sharp magnetic DWs, via a magnetostrictive mechanism~\cite{PhysRevLett.125.067602,PhysRevLett.125.067602,PhysRevB.96.104431,C9TC02501D,doi:10.1063/1.4917560}, neglecting spin-orbit coupling (SOC). Such a description is of limited use for DWs in real materials, that can span up to hundreds of nanometers.
A more sophisticated treatment calls for multiscale methods enabling an accurate description of magnetoelectric effects at different length scales, with realistic, materials specific parameters, determined from first principles, electronic structure calculations. Such studies have recently appeared in literature, using complex magnetoelectric models that often lack careful and systematic validation. In some cases the studies show conspicuous discrepancies, as is the case, for example, for the 2D magnet CrI$_3$.~\cite{Behera_CrI3_Efield,Ghosh_CrI3_Efield,PhysRevMaterials.4.094004}

Since the discovery of long-range magnetic order in 2D monolayers of Van Der Waals materials, such as CrI$_3$~\cite{Huang2017,CGT_gong,mcguire,Burch2018}, massive research efforts have led to exciting developments, including experimental realisation of 2D monolayer magnets exhibiting multiferroicity~\cite{Song2022} and Skyrmions~\cite{Han2019}. 
These ultrathin magnets are interesting in the context of magnetoelectric effects; a moderate voltage is sufficient to produce a huge electric field in the perpendicular direction, and even metallic magnets may be possibly manipulated using such fields when the penetration length exceeds the thickness. 
Moreover, their high flexibility and tendency to ripple raises a number of fundamental questions about the possible interplay of magnetoelectric effects and layer curvature.
The recent surge of interest in flexoelectricity (i.e., the linear coupling between strain gradients and electric polarization~\cite{Flexoel_Zubkoetal,springolo2021,springolo2023inplane}) and flexomagnetism (defined as a modification of the spin parameters due to curvature~\cite{hertel}) thus provides additional motivation to explore magnetoelectric effects in the same materials, especially in 2D magnets such as CrI$_3$~\cite{our_PRL}.
And indeed, recent works have revealed intriguing similarities between the phenomenology of applied electric fields and flexural deformations: for example, both lead to the appearance of an induced Dzyaloshinskii-Moriya interaction (DMI)~\cite{Sheka_2015,Streubel_2016,Sheka2021,PhysRevMaterials.4.094004}, which is a universal feature of systems with broken space inversion symmetry~\cite{DZYALOSHINSKY1958241,PhysRev.120.91}, and an important ingredient in the stabilization of topological structures.

Here, we develop a multiscale approach, described in Sec.~\ref{sec.meth}, considering atomistic and continuum models of magnetoelectricity, with all parameters determined using first principles density functional theory (DFT) calculations. We carefully validate these models against direct DFT calculations of other complex spin states that differ from those used to build the model. 
We also show how the magnetoelectric coupling parameters correspond precisely to the linear terms of the electric field induced DMI, in both the atomistic and continuum forms. Next, we apply the models to study the electric polarization of spin spirals in Sec.~\ref{sec.spirals} and magnetic topological solitons in the form of DWs and Skyrmions in Sec.~\ref{sec.DW}-\ref{sec.Skyrmion}, and assess the possibility of magnetoelectric engineering of these topological solitons. Interestingly, while most studies of electric field effects on 2D magnetism has focused on perpendicular electric fields~\cite{Behera_CrI3_Efield,PhysRevB.108.134430}, we find that the magnetoelectric coupling in CrI$_3$ is more than an order of magnitude stronger for in-plane electric fields. The continuum model provides a particularly simple description in terms of only three parameters, giving useful insights into the magnetoelectric phenomena, for example showing that Skyrmions carry out-of-plane electric dipole moments, %
while anti-Skyrmions carry in-plane dipole moments.

\section{Methods and Models}\label{sec.meth}

Early discussions of the electric polarization induced by non-collinear spin configurations and spin spirals include  phenomenological descriptions, as in earlier Ref.~[\onlinecite{inhom_magnetoel}] or in Ref. [\onlinecite{PhysRevLett.96.067601}], and microscopic mechanisms, such as the purely electronic one proposed by Katsura, Nagaosa and Balatsky (KNB) relating spin currents and electric polarization~\cite{KNB}. The KNB model is essentially a dimer model comprising two magnetic atoms in a cubic environment, predicting a polarization contribution in direction $\hat{\mathbf{e}}_{12} \times (\hat{\mathbf{e}}_{1} \times \hat{\mathbf{e}}_{2})$ for two spins parallel to $\hat{\mathbf{e}}_{1}$ and $\hat{\mathbf{e}}_{2}$, respectively, and separated in the direction $\hat{\mathbf{e}}_{12}$. Despite its physical insight, the KNB result is limited to cases described by the model itself, which is a very specific one, failing, e.g., in describing the spin-driven ferroelectricity of triangular-lattice multiferroics of the delafossite family\cite{SekiPRL2008}. 
However, the underlying idea of decomposing spin-induced polarization in spin-dimer contributions have been later pursued\cite{kaplan_prb2011, PhysRevLett.107.157202}, providing a generalization of the KNB model in the form of a pairwise, interatomic, coupling tensor. {The latter} has successfully described the electric polarization induced by complex spin textures in low-symmetry crystal classes\cite{gKNB_theo_YangPRL2012,gKNB_theoII_LuPRL2012,gKNB_expI_HwangPRL2012,gKNB_expII_SHARMA2015,gKNB_exp3_ChaiPRB2018,gKNB_exp4_TangPRB2021}, including 2D multiferroics~\cite{Song2022}. This is the atomistic model which we use and describe further in Sec.~\ref{sec.gKNB}.

Considering a phenomenological continuum model in cubic symmetry, Mostovoy~\cite{PhysRevLett.96.067601} found a similar result as Katsura et al., showing, for example, that a helix has zero polarization while a cycloid carries a polarization parallel to $\hat{\mathbf{e}}_3 \times \mathbf{q}$, where $\hat{\mathbf{e}}_3$ is the spin rotation axis and $\mathbf{q}$ the propagation vector. Here we adopt a more general continuum form of the inhomogeneous magnetoelectric effect, which was introduced by Baryakhtar et al.~\cite{inhom_magnetoel} without assuming a specific crystalline symmetry (see Sec.~\ref{sec.contmod}).

In the following two sections we introduce the atomistic and continuum models of magnetoelectricity used in this work, respectively, and determine their mutual connections. Next, in Sec.~\ref{sec.DFT} we describe the DFT calculations that we use to determine the model parameters from {first principles}. Finally, in Sec.~\ref{sec.param} we present our calculated values, and  in Sec.~\ref{sec.valid} we validate them on representative spin structures that differ from those used to construct our model.

\subsection{Atomistic Model}\label{sec.gKNB}

Within an atomistic approach, the macroscopic polarization is decomposed in atomic and pairwise contributions\cite{kaplan_prb2011, PhysRevLett.107.157202}, as detailed in Appendix \ref{sec:app_macroP}. In centrosymmetric systems, the magnetically induced polarization arising from non-collinearity of spins in a helimagnetic configuration takes the form of a generalized spin-current (gKNB) coupling:
\begin{equation}\label{eq.P_Mtensor}
    \mathbf{P} = \frac{1}{2\Omega}\sum_{i,j} \bm{\mathcal{M}}^{ij} (\mathbf{S}_i \times \mathbf{S}_j),
\end{equation}
where $\Omega$ denotes the unit cell volume (area) in the 3D (2D) case, the $i$ index denotes all lattice sites in the cell, $j$ runs over all neighbors for each given site (typically restricted to few nearest-neighbour shells), while the prefactor $1/2$ is introduced to count each bond contribution only once. We consider normalized spins. Note the antisymmetric character of the interaction in the two participating spin components; indeed, it can be shown that only the cross product of the spins contribute to the macroscopic polarization induced by spiral magnets (see Appendix \ref{sec:app_macroP}). 
The 3$\times$3 matrix $\bm{\mathcal{M}}^{ij}$ generalizes the standard KNB result, describing the magnetoelectric (spin-dipole) coupling. In the most general case, it is uniquely determined by 9 coefficients for each bond. Crystalline symmetries introduce constraints on the form of the coupling tensor and its allowed nonzero coefficients, both in deriving the transformation rules relating $\bm{\mathcal{M}}^{ij}$ of different symmetry-equivalent bonds and in determining its symmetry-allowed form  for each bond, that will depend on the subset of symmetry operation of the bonding vector (Appendix \ref{sec:app_symmM}).

In a first-principles context, the magnetoelectric interaction matrix $\bm{\mathcal{M}}^{ij}$ can be calculated following the procedure outlined in Ref.~[\onlinecite{PhysRevLett.107.157202}]. 
{In particular}, each column of the matrix for a given pair of spins is evaluated from the linear combinations of polarizations calculated for a set of selected non-collinear spin-dimer states at fixed atomic positions. 
The last assumption implies that only the electronic contributions to the spin-induced polarization are taken into account, and they will be our {exclusive} focus in the following. 
However, the analysis can be generalized to include also the spin-order-induced ionic contribution to macroscopic polarization, following, e.g., Ref.~[\onlinecite{PhysRevB.88.054404}]. As mentioned above, the approach has been successfully applied to describe the spin-driven polarization of several low-symmetry materials, such as triangular lattice multiferroics\cite{Song2022,gKNB_theo_YangPRL2012,gKNB_theoII_LuPRL2012,gKNB_expI_HwangPRL2012,gKNB_expII_SHARMA2015,gKNB_exp3_ChaiPRB2018,gKNB_exp4_TangPRB2021}.

The expression in Eq.~\ref{eq.P_Mtensor} takes on a strikingly similar form as the atomistic Dzyaloshinskii-Moriya interaction (DMI). Comparing the electric dipole energy $E = \bm{\mathcal{E}} \cdot \mathbf{p}$ of a homogeneous electric field $\bm{\mathcal{E}}$ coupled to an electric dipole $\mathbf{p}$, we can immediately extract the electric field induced DMIs. 
The DMI energy contribution of a pair of atomic spins is~\cite{PhysRev.120.91}
\begin{equation}
    E_{ij} =\mathbf{D}^{ij} \cdot (\mathbf{S}_i \times \mathbf{S}_j),
\end{equation} 
while the electric dipole moment of the same spin pair is $\mathbf{p}_{ij} = \bm{\mathcal{M}}^{ij} (\mathbf{S}_i \times \mathbf{S}_j)$, with energy $E_{ij} = -\bm{\mathcal{E}} \cdot \mathbf{p}_{ij}$ 
Equating these expressions for $E_{ij}$ gives the electric field induced atomistic DMI 
\begin{equation}\label{eq.atomisticDMI}
    \mathbf{D}^{ij} (\bm{\mathcal{E}} ) = -(\bm{\mathcal{M}}^{ij})^T \bm{\mathcal{E}} \quad \text{or} \quad \mathcal{M}^{ij}_{\alpha\beta} = -\frac{\partial D^{ij}_\beta}{\partial \mathcal{E}_\alpha}.
\end{equation} 
As shown in Appendix \ref{sec:app_symmM}, it is straightforward to check that Moriya's rules for the Dzyaloshinskii vector $\bm D(\bm{\mathcal{E}})$ are easily recovered using the symmetry constraints on the local coupling tensor $\bm{\mathcal{M}}$ and identifying the symmetry operations compatible with the applied field $\bm{\mathcal{E}}$.
Thus, the magnetoelectric coupling tensors are equivalent to the electric-field induced DMIs, and can be used, for example, to simulate an applied electric field in atomistic spin dynamics simulations.
Although a close connection between magnetoelectricity and DMI was pointed out already by Sergienko and Dagotto~\cite{PhysRevB.73.094434}, to the best of our knowledge, the direct relation in Eq.~\ref{eq.atomisticDMI} was overlooked in earlier works; we regard it as one of our main formal results. 

\subsection{Continuum Model}\label{sec.contmod}

Baryakhtar et al.\cite{inhom_magnetoel} introduced a phenomenological continuum description of the inhomogenous magnetoelectric effect, where the \emph{local} polarization is given by $P_\alpha = f_{\alpha\beta,\delta\gamma}M_\delta \partial_\beta M_\gamma$. (${\bf M}$ is the ferromagnetic order parameter and $\partial_\beta$ is a short-hand symbol for the derivative along the Cartesian coordinate $r_\beta$.) This is indeed the most general form of the coupling between an arbitrary (pseudo)vector field and ${\bf P}$~\cite{PhysRevLett.132.146801}: it can be trivially generalized to the antiferromagnetic order parameter as well.
Nevertheless, one can show via an integration by parts that only the antisymmetric components of the tensor in the Cartesian indices $\alpha \beta$ contribute to the macroscopic polarization~\cite{PhysRevLett.96.067601}.
To reflect this fact, we write
\begin{equation}\label{eq.contmod}
P_\alpha = \tilde{f}_{\alpha\beta,\delta\gamma}\left( M_\delta \partial_\beta M_\gamma - M_\beta \partial_\beta M_\alpha\right), 
\end{equation}
where we have introduced the anti-symmetric tensor $\tilde{f}_{\alpha\beta,\delta\gamma} = \frac{1}{2}( f_{\alpha\beta,\delta\gamma} - f_{\alpha\beta,\gamma\delta} )$ (see Appendix~\ref{app.contmodel}).

In the following we show that $\tilde{f}$ is directly related to the coefficients of the atomistic model, and establish a closed formula that yields the former as a function of the latter.
The values in $\tilde{f}_{\alpha\beta,\gamma\delta}$ have a clear physical interpretation as the linear part in $q_\beta$ of the polarization component $\alpha$,  arising for a spin spiral propagating along Cartesian direction $\beta$, with the spins rotating in the plane spanned by Cartesian directions $\gamma$, $\delta$. 
Considering Eq.~\ref{eq.P_Mtensor} in the small-${\bf q}$ limit,
each type of spin spiral (there are three possible planes of rotation for spins, propagating along an arbitrary Cartesian direction), yields three conditions for  
any given component $P_\alpha$, leading to the 27 (3D) or 18 (2D) conditions needed to fully determine $\tilde{f}_{\alpha\beta,\gamma\delta}$ in terms of $\bm{\mathcal{M}}^{ij}$. These can be compactly written as
\begin{equation}\label{eq.atomistic_cont}
  \tilde{f}_{\alpha\beta,\gamma\delta}= \frac{1}{\Omega}\sum_{i,j,\lambda} \epsilon_{\gamma\delta \lambda} r_{\beta}^{ij} \mathcal{M}^{ij}_{\alpha \lambda}  ,
\end{equation}
where $\epsilon_{\alpha \beta \gamma}$ is the anti-symmetric tensor and $r_{\beta}^{ij}$ is the component $\beta$ of the vector $\mathbf{r}^{ij}$ connecting spins $i$ and $j$.

Next, we show that the relation to the DMI energy also holds in the continuum case, where the DMI energy density is~\cite{PhysRevB.100.214406}
\begin{equation}\label{eq.contDMI}
E = D_{\alpha\beta\gamma} \left( M_\alpha \frac{\partial M_\beta}{\partial r_\gamma} - M_\beta \frac{\partial M_\alpha}{\partial r_\gamma} \right). 
\end{equation}
$D_{\alpha\beta\gamma}$ vanishes identically in presence of space-inversion (SI) symmetry, but an external electric field breaks SI and allows for a nonzero ${\bf D}$.
Comparing the energy density in Eq.~\ref{eq.contDMI} with that of an electric field $\mathbf{\mathcal{E}}$ coupled to the polarization in Eq.~\ref{eq.contmod}, we find 
\begin{equation}\label{eq.cont_DMI_rel}
\tilde{f}_{\alpha\beta,\gamma\delta} = - \frac{\partial D_{\gamma\delta \beta}}{\partial \mathcal{E}_\alpha}.
\end{equation}  
This result is the continuum counterpart of Eq.~\ref{eq.atomisticDMI}, relating the microscopic DMI and the spin-polarization coupling coefficients, and 
allows the effect of electric fields to be studied in micromagnetic models of modulated spiral orders. As such, Eq.~\ref{eq.atomistic_cont} can be considered an alternative form of the usual relation between the atomistic and continuum DMI terms, available e.g. in Ref.~[\onlinecite{PhysRevB.100.214406}].

\subsection{Electrical boundary conditions}

Since the couplings discussed here involve the macroscopic electrical polarization, the electrical boundary conditions (EBC) need to be specified in their definition.
For a 3D crystal, the most natural choice is to assume short-circuit EBC, consistent with the extended nature of the system in all directions. 
This choice is automatically enforced by the assumption of periodicity in standard first-principles codes, and does not require any special treatment.

In the case of an isolated 2D layer, which will be our main focus in the remainder of this work, the most natural choice for the EBC consists in assuming free boundary conditions.~\cite{PhysRevX.11.041027}
This requires some care, as it requires in principle a truncated Coulomb kernel~\cite{PhysRevB.73.233103,PhysRevB.73.205119,PhysRevB.96.075448,PhysRevX.11.041027}, which at the zone center (relevant to the macroscopic polarization targeted in this work) boils down to applying a dipole correction in the vacuum layer. 
In this work we find it more practical to perform all our calculation within the standard 3D electrostatic kernel, and correct a posteriori for the spurious unphysical interaction between images.
More specifically, we extrapolate the $z$-component of the polarization to the limit of infinite vacuum using the expression 
\begin{equation}\label{eq.P_c_fit}
  P_z(c) = \frac{P_z(c \rightarrow\infty)}{1-\frac{\alpha_\perp}{\epsilon_0 c}},  
\end{equation} 
where $c$ is the distance between neighboring monolayers and $\alpha_\perp$ is
the out-of-plane polarizability of the layer at frozen spins~\cite{PhysRevX.11.041027}. 

\subsection{Monolayer CrI$_3$}\label{sec.CrI3}

\begin{figure} 
	\centering
 	\includegraphics[width=0.45\textwidth]{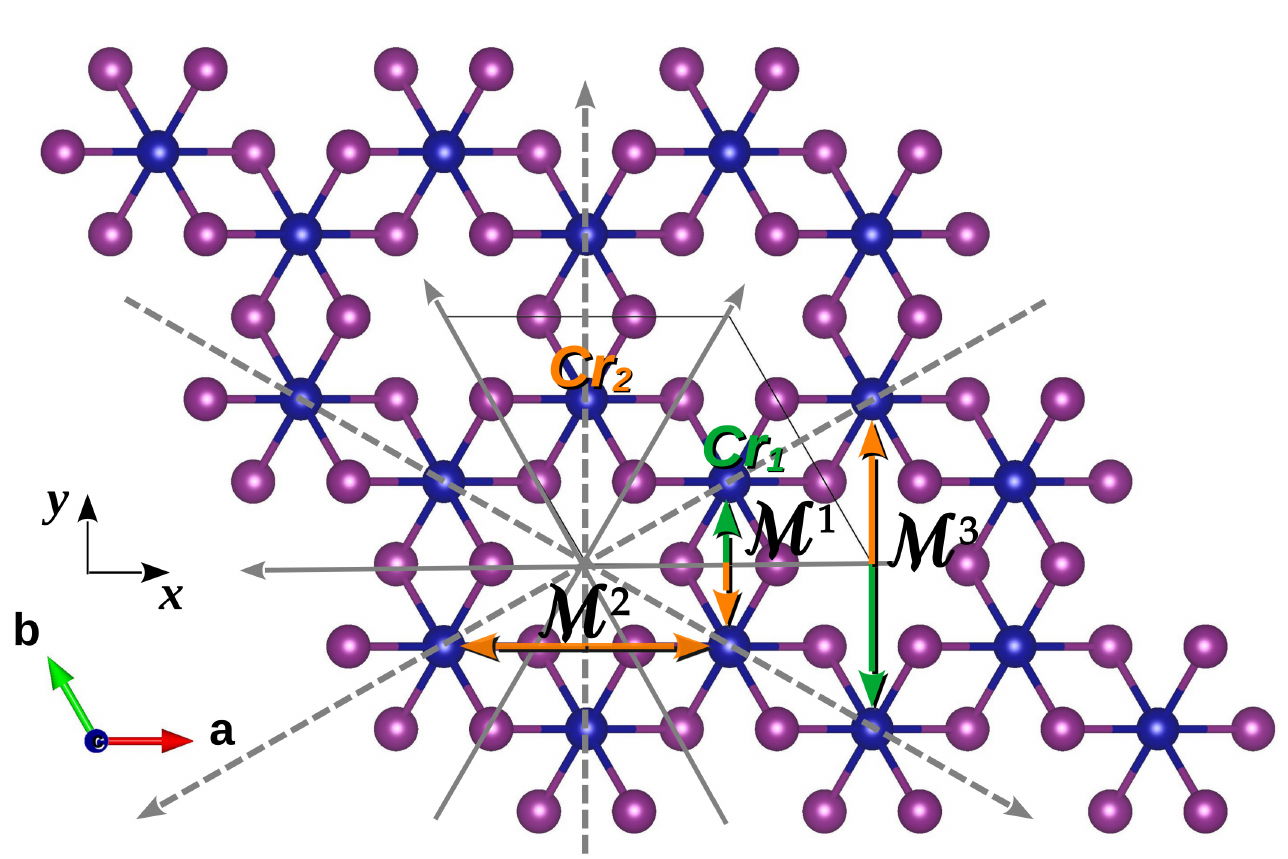} \\
	\caption{Crystal structure of the CrI$_3$ monolayer, highlighting the unit cell, the adopted Cartesian reference frame as well as symmetry axes and planes. Dashed lines represent the in-plane twofold rotation axes, solid lines the vertical mirror planes. The threefold rotation/rotoreflection axis is perpendicular to the monolayer plane. Inversion-partner Cr sites  are highlighted within the unit cell. Two-headed arrows show the 1st, 2nd and 3rd NN bonds whose coupling tensors $\bm{\mathcal{M}}_i$ have been evaluated; the presence of an inversion center in the bond is highlighted by bicoloured arrows.}
	\label{fig.CrI3_struct}
\end{figure}

Here we briefly summarize the crystal structure and general properties of CrI$_3$ monolayer, which we analyze in the remainder of the work as an illustrative demonstration of our approach.
Fig.~\ref{fig.CrI3_struct} displays the crystal structure of monolayer CrI$_3$ with space-group symmetry P$\Bar{3}1m$ and crystallographic point group $D_{3d}$, comprising inversion, threefold rotation/rotoreflection axis orthogonal to the monolayer, three twofold in-plane rotational axes and three vertical mirrors highlighted by dashed and solid lines, respectively. Cr atoms occupy the $2c$ Wyckoff position, with local $D_3$ symmetry, the two Wyckoff partners being related by inversion symmetry. The lattice constant is $a=6.71~\mathrm{\AA}$, corresponding to a Cr-Cr distance of $\frac{a}{\sqrt{3}}=3.87~\mathrm{\AA}$. Monolayers have been experimentally exfoliated and order ferromagnetically, with out-of-plane magnetization, below 45~K~\cite{Huang2017}. 
The magnetism of a material with uniaxial anisotropy is described by an energy 
\begin{eqnarray}\label{eq.mag_energy}
   E = A\int \left( \left[ \nabla \mathbf{M} \right]^2 - \frac{1}{\xi^2} M_z^2 \right) d^2r
\end{eqnarray}
 with energy scale given by the spin stiffness $A$,  and characteristic length scale $\xi = \sqrt{\frac{A}{K}}$, where $K$ is the magnetic anisotropy. The 1st nearest-neighbor (NN) Cr-Cr bond has an inversion center whereby the corresponding interatomic DMI is zero, just like the continuum DMI, while the 2nd NN bond, which lacks inversion center, has a DMI of around 60~$\mathrm{\mu eV}$ ~\cite{PhysRevB.102.115162} with an associated appearance of topological magnons~\cite{PhysRevX.8.041028}. 
(See Fig. \ref{fig.CrI3_struct} for an illustration of the bonds.)

By taking into account the above symmetries, the magnetoelectric tensors associated to the the bonds displayed in Fig. \ref{fig.CrI3_struct} take the form (see Appendix \ref{sec:app_symmM} for details)
\begin{align}\label{eq:M1_M3}
\bm{\mathcal{M}}^1 & = \begin{bmatrix}
\mathcal{M}^1_{11}   & 0          &  \mathcal{M}^1_{13} \\
0            & \mathcal{M}^1_{22} &  0          \\
\mathcal{M}^1_{31}   &  0         &  \mathcal{M}^1_{33}
\end{bmatrix}, 
\end{align}
\begin{align}\label{eq:M2}
\bm{\mathcal{M}}^2 & = \begin{bmatrix}
0         & \mathcal{M}_2^{12}    &  0        \\
\mathcal{M}^2_{21}  & 0           &  \mathcal{M}^2_{23} \\
0         & \mathcal{M}^2_{32}    &  0
\end{bmatrix}.
\end{align}
The 3rd NN tensor $\bm{\mathcal{M}}^3$ takes the same form as $\bm{\mathcal{M}}^1$.
In principle, the local site- and bond-symmetries of CrI$_3$ would allow for additional (symmetric) intra- and inter-site contributions to $\mathbf{P}$ (see Appendices A,B). Besides being numerically small, these terms do not contribute to the macroscopic electric polarization of spin spirals nor general spatially slowly varying spin textures, and they will be neglected in the following.

As described in Appendix~\ref{app.contmodel}, the most general tensor of the continuum model $f_{ik,\alpha\beta}$ is reduced to have 9 independent non-zero components in a CrI$_3$ monolayer, but only three remain in the anti-symmetric part, $\tilde{f}_{12,12}$, $\tilde{f}_{11,31} $ and $\tilde{f}_{31,31}$. The component $\tilde{f}_{12,12}$ corresponds to the in-plane polarization of a cycloid with spins in the plane, and is mainly important in systems with in-plane magnetization ($K<0$), while it will be of less importance in CrI$_3$ with perpendicular magnetization $K>0$. The other two components, which we label $\tilde{f}_\mathrm{ip} \equiv \tilde{f}_{11,31} $ and $\tilde{f}_z \equiv \tilde{f}_{31,31}$, correspond to the in-plane and out-of-plane components of polarization for a cycloid with out-of-plane spins. 
In high symmetry materials (point groups $O_h$, $T_d$ and $O$, see Appendix~\ref{app.contmodel} ), $\tilde{f}_\mathrm{ip}=0$ while $\tilde{f}_z$ and $\tilde{f}_{12,12}$ are equal and non-zero, as in the model studied by Mostovoy~\cite{PhysRevLett.96.067601}, consistent with the KNB model~\cite{KNB}. The KNB model corresponds to the only non-zero elements of $\mathcal{M}^{ij}_{\alpha\beta}$ being $\mathcal{M}^1_{31} = - \mathcal{M}^1_{13}$, $\mathcal{M}^2_{32} = - \mathcal{M}^2_{23}$ and $\mathcal{M}^3_{31} = - \mathcal{M}^3_{13}$, that is precisely those which determine $\tilde{f}_z$ and $\tilde{f}_{12,12}$, which would be equal under such conditions (see Eq.~\ref{eq.fz_M}-\ref{eq.f1212}).
The polarization due to $\tilde{f}_\mathrm{ip}$ can be considered a magnetoelectric analogue to the unconventional, in-plane flexoelectric response studied by Springolo et al.~\cite{springolo2023inplane}. Note, however, that while the flexoelectric tensor is also a fourth rank tensor, but symmetric in the last two indices. 
The magnetoelectric response, which we focus on here, instead is given by a rank four tensor which is anti-symmetric in the last two indices.

\section{First-principles calculations}

\subsection{Computational parameters}\label{sec.DFT}

In order to determine all parameters of the atomistic model in Sec.~\ref{sec.gKNB}, which are then used to determine the continuum parameters, we use density functional theory (DFT) calculations. Calculations are performed using the projector augmented wave (PAW)~\cite{PhysRevB.50.17953,PhysRevB.59.1758} method, as implemented in VASP~\cite{KRESSE199615,PhysRevB.49.14251,PhysRevB.47.558}, with settings in accordance with earlier work~\cite{Xu2018,our_PRL}. The local density approximation (LDA) is used for the exchange-correlation together with an additional Coulomb repulsion~\cite{PhysRevB.57.1505} of $U=0.5~\mathrm{eV}$ on  Cr d-states. %
Calculations for an isolated monolayer are performed within periodic boundary conditions, with (at least) {15~\AA} of vacuum separating the repeated monolayer images. 
Constrained, non-collinear magnetic calculations, with SOC, were done using a penalty energy for spins deviating from the desired configuration~\cite{PhysRevB.62.11556,PhysRevB.91.054420}. We checked that the results are insensitive to the value of the penalty energy. 
The calculations of the magnetoelectric tensor described in Sec.~\ref{sec.gKNB}  were done in a $3\times 3 \times 1$ supercell with a $6\times 6 \times 1$ $k$-point grid. 
In Sec.~\ref{sec.valid} we also present validations of the models using direct DFT calculations for the polarization of spin spirals. These refer to spin spirals with wavevector $q=\frac{2\pi}{L}$, performed in a supercell of dimension $a\times L$ or $\sqrt{3}a \times L$, for propagation in the $x$ or $y$-direction, respectively, where $a=6.71~\mathrm{\AA}$ is the lattice constant.  The electric polarization is evaluated with the Berry phase formalism~\cite{PhysRevB.47.1651} in all cases. 

\begin{figure}[h!]
	\centering
	\includegraphics[width=0.45\textwidth]{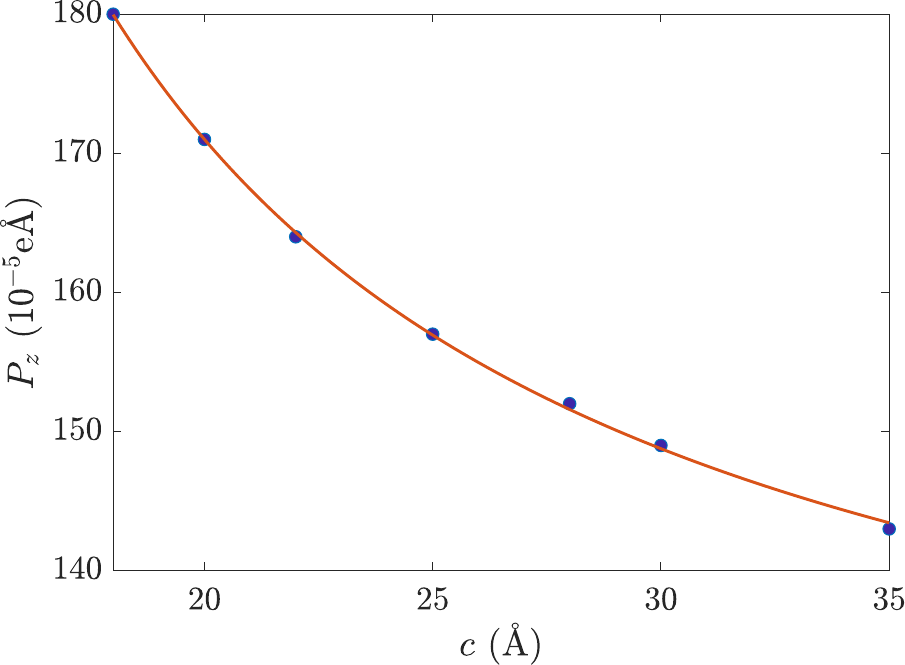}
	\caption{$P_z$ calculated as a function of the distance $c$ between adjacent monolayer periodic images, for one spin configuration, including a fitting of the form $P_z(c) = \frac{P_z(c \rightarrow\infty)}{1-\frac{\alpha_\perp}{\epsilon_0 c}}$, leading to $\alpha_\perp=5.5\cdot 10^{-21}~\mathrm{F}$. 
    }
	\label{fig.Pz_of_c}
\end{figure}

Regarding the electrical boundary conditions, by varying the interlayer distance $c$ in the range 17.5~\AA~to 35~\AA~we find an essentially perfect fit to Eq.~\ref{eq.P_c_fit} with
$\alpha_\perp=5.5\cdot 10^{-21}~\mathrm{F}$, as shown in Fig.~\ref{fig.Pz_of_c}.

\subsection{Results for the calculated physical constants}\label{sec.param}

The magnetic parameters of Eq.~\ref{eq.mag_energy} are taken from an earlier calculation on the same material~\cite{our_PRL}, where we found $K = 0.75~\mathrm{meV / Cr}$ and $A = 48~\mathrm{meV\AA}^2 / \mathrm{Cr}$, corresponding to $\xi = \sqrt{A/K} = 8~\mathrm{\AA}$. %

We calculated $\bm{\mathcal{M}}^{ij}$ up to fourth nearest neighbour (NN), following the procedure outlined in Ref.~[\onlinecite{PhysRevLett.107.157202}]. The 2$^\mathrm{nd}$ NN needs special care as the bond lacks inversion center, leading to non-zero symmetric elements. Hence, the full intersite tensor in Eq.~\ref{eq.intersiteP} needs to be calculated and anti-symmetrized. The values for 1$^\mathrm{st}$, 2$^\mathrm{nd}$ and 3$^\mathrm{rd}$ NNs are (in units of $10^{-5}\mathrm{e\AA}$)
\begin{align} \label{Eq.M1num}
\bm{\mathcal{M}}^1 & = 
\begin{bmatrix}
-79  & 0 &  -98 \\
         0  & -20 &       0 \\
  -59   &      0  &  6
\end{bmatrix}   \\ 
\bm{\mathcal{M}}^2 & =
\begin{bmatrix}
0 & 273 &  0 \\
61 & 0 &  112 \\
0   &  11  &  0
\end{bmatrix}  \\
\bm{\mathcal{M}}^3 & = 
\begin{bmatrix}
109 &  0     &  -33 \\
0   &  55    &   0    \\
0   &  0     &   0
\end{bmatrix}. 
\label{Eq.M3num}
\end{align}
Some of the elements that are zero by symmetry numerically come out as order of magnitude $1 \cdot 10^{-5}\mathrm{e\AA}$, indicating the numerical accuracy of the calculations. The $4^\mathrm{th}$ NN is smaller than the former three, with its largest element being $17~\cdot 10^{-5}\mathrm{e\AA}$, and is neglected in the following. $\bm{\mathcal{M}}^2$ has larger components than $\bm{\mathcal{M}}^1$ and additionally has a higher coordination of six, compared to three for $\bm{\mathcal{M}}^1$. Consequently, we expect $\bm{\mathcal{M}}^2$  to be more important than $\bm{\mathcal{M}}^1$ in describing the magnetoelectric polarization in CrI$_3$. 
Interestingly, the 3$^\mathrm{rd}$ NN magnetoelectric interaction tensor $\bm{\mathcal{M}}^3$ also has larger elements than those of $\bm{\mathcal{M}}^1$. 
Thus, all of the three NN interactions can be important, and as will be seen in the next section, including the three NN ME tensors is necessary and sufficient to accurately describe a variety of complex spin structures. %
We also determined the intrasite tensor (first term of Eq.~\ref{eq:general_P}) and symmetric intersite part for the 2$^\mathrm{nd}$ NN, and confirmed these also to be small compared to Eq.~\ref{Eq.M1num}-\ref{Eq.M3num}. Thanks to the global space inversion (SI) symmetry of the crystal, neither of the latter would contribute to the macroscopic polarization induced by helimagnetic ordering.

Next, the parameters of the continuum model are determined from those of the discrete model, using the relations from Eq.~\ref{eq.atomistic_cont}, as listed below. 
\begin{widetext}
\begin{align}
    \tilde{f}_\mathrm{ip} \equiv \tilde{f}_{11,31} & = \frac{1}{a} \left( \mathcal{M}^{1}_{22} - \mathcal{M}^{1}_{11} + 2 \sqrt{3}(\mathcal{M}^{2}_{21} + \mathcal{M}^{2}_{12}) + 2(\mathcal{M}^{3}_{11} - \mathcal{M}^{3}_{22} ) \right) = 198\cdot 10^{-5}e \\ 
    \tilde{f}_{z} \equiv \tilde{f}_{31,31} & = \frac{2}{a} \left( \mathcal{M}^{1}_{31} + 2\sqrt{3} \mathcal{M}^{2}_{32} - 2 \mathcal{M}^3_{31} \right) = -8\cdot 10^{-5}e \label{eq.fz_M} \\ 
    \tilde{f}_{12,12} & = -\frac{2}{a} \left( \mathcal{M}^1_{13} + 2\sqrt{3} \mathcal{M}^2_{23} - 2 \mathcal{M}^3_{13}  \right) = -106\cdot 10^{-5}e \label{eq.f1212}
\end{align}
\end{widetext}

Given the relations in Eq.~\ref{eq.atomisticDMI} and \ref{eq.cont_DMI_rel}, the numerical values listed above are the linear coefficients for the electric field induced DMI. 
The DMI induced by finite electric fields has been calculated from first principles for monolayer CrI$_3$ in several works~\cite{PhysRevB.97.054416,PhysRevMaterials.4.094004,Behera_CrI3_Efield,Ghosh_CrI3_Efield}. 
There are, however, discussions~\cite{Behera_CrI3_Efield,Ghosh_CrI3_Efield} and notable discrepancies between the data available in literature. In particular, calculations in the generalized gradient approximation (GGA) produce results varying over 2 orders of magnitude and sign~\cite{PhysRevB.97.054416,Behera_CrI3_Efield,PhysRevMaterials.4.094004}. Meanwhile, our result for the 1$^\mathrm{st}$ NN DMI, $D^1_z/\mathcal{E}_z = - \mathcal{M}^1_{33} =  -6\cdot 10^{-5}~\mathrm{e\AA}$ is in reasonable agreement with the LDA calculation of Ghosh et al.~\cite{Ghosh_CrI3_Efield} giving a value of $D^1_z/\mathcal{E}_z = -13.5\cdot 10^{-5}~\mathrm{e\AA}$, even though our respective approaches differ: while Ghosh et al. calculated the DMI from energy differences under a finite electric field, and under assumption that further neighbour DMI is neglibible, while we calculated electric polarizations arising from non-collinear spin arrangements.

Earlier works have often focused on the out-of-plane DMI component $D_z$ for a perpendicular electric field $\mathcal{E}_z$ and the first nearest neighbour bond~\cite{PhysRevB.97.054416,Behera_CrI3_Efield,Ghosh_CrI3_Efield}. Hence, it is also interesting to note that the component $\mathcal{M}^1_{33}$, which this corresponds to in the magnetoelectric interaction tensors, is relatively small. In fact, in-plane electric fields will be much more effective in inducing DMI, and as we have seen, $2^\mathrm{nd}$ and $3^\mathrm{rd}$ NN interactions are at least as, or more, important than the first one. Finally, we point out that as the magnetoelectric tensors provide a complete description of the DMI induced by an arbitrary electric field, they provide an excellent starting point for further studies of electric field effects on magnetism, e.g. using atomistic spin dynamics simulations~\cite{Eriksson2017}, allowing dynamics and finite temperatures to be investigated. To do that accurately one should include also lattice-mediated effects, in addition to the electronic ones considered here. 
Ghosh et al.~\cite{Ghosh_CrI3_Efield} argued that the effect of electric field on the crystal structure is small, but the impact on the DMI parameters remains to be addressed.

Recently, the curvature induced continuum DMI in CrI$_3$ was calculated to $\frac{D_{zyy}}{\kappa} = 3.3~\mathrm{meV}$~\cite{our_PRL}, compared to $\frac{D_{zyy}}{\mathcal{E}_z} = 181\cdot 10^{-5}~\mathrm{e}$ found here. The ratio of these values is $\frac{\mathcal{E}_z}{\kappa} = 1.8~\mathrm{V}$, which implies that an electric field of 1.8~mV/\AA~ induces a similar DMI as a curvature of $10^{-3} \mathrm{\AA}^{-1}$.  In Ref~[\onlinecite{our_PRL}] we predicted that a curvature of $0.07~\mathrm{\AA}^{-1}$ stabilizes a spin cycloid with lower energy than the FM state. In the following, we will find the analogous result of electric field induced spin cycloids with consistent order of magnitude of the critical field.

\subsection{Validation}\label{sec.valid}

\begin{figure} 
	\centering
	\includegraphics[width=0.40\textwidth]{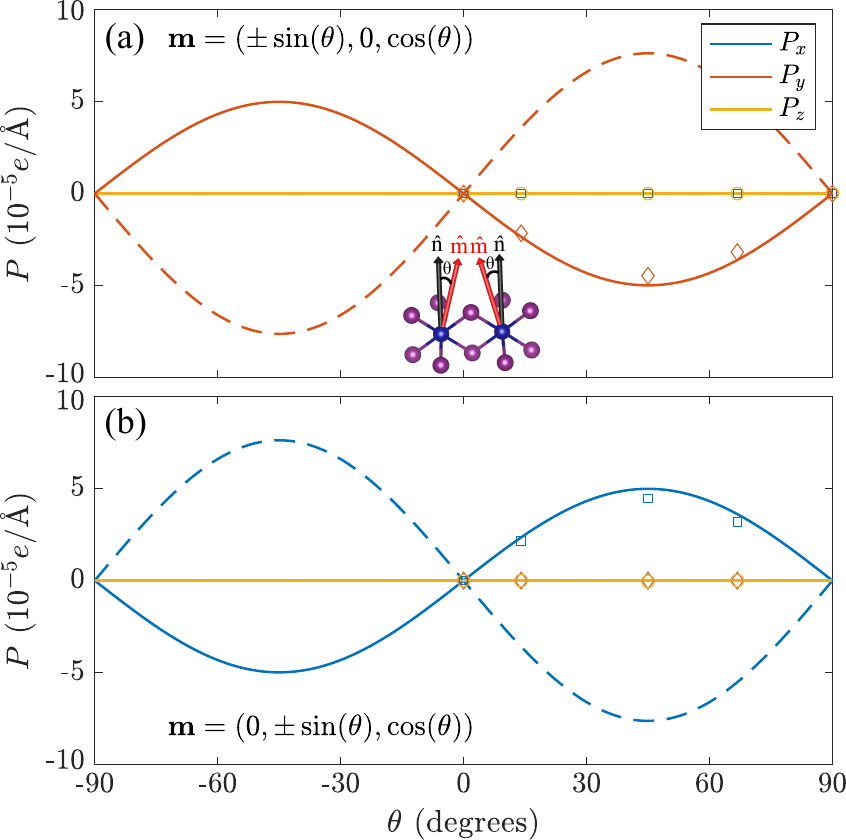} \\
	\caption{ Electric polarization as a function of tilt angle $\theta$, with neighbouring Cr spins tilted in opposite $x$ or $y$-directions, in bottom and top panels respectively. Markers show the result of direct DFT calculations, while the lines show the results of Eq.~\ref{eq.P_Mtensor} using only 1st NN (dashed lines) or up to 3rd NN (solid lines) coupling. 
 }
	\label{fig.CrI3_P_AFM}
\end{figure}

To validate the gKNB model and its computed parameters, we compare the predictions of the model to direct DFT calculations of the electric polarization for a representative non-collinear spin structure that differs from those used to numerically determine $\bm{\mathcal{M}}^{ij}$. 
In particular, we consider a tilt of the spins corresponding to a $q=0$ optical magnon mode. 
This means that every spin is tilted by an angle $\pm\theta$ along some in-plane direction $\hat{s}=\left(\cos\alpha , \sin\alpha, 0 \right)$, in such a way that the tilt angle alternates in sign between $1^\mathrm{st}$ NN sites. 
Fig.~\ref{fig.CrI3_P_AFM} shows the resulting electric polarization as a function of tilt angle $\theta$ for two distinct choices of $\hat{s}$: in (a) $\hat{s}=(1,0,0)$ is aligned with the $\hat{x}$ Cartesian direction, while $\hat{s}=(0,1,0)=\hat{y}$ in (b).
The solid lines, showing the results of the full atomistic model (Eq.~\ref{eq.P_Mtensor}), are
in excellent agreement with the reslts of direct DFT calculations (symbols).
Conversely, by truncating the model to 1$^\mathrm{st}$ interactions (dashed curves), we obtain a marked discrepancy, both in magnitude and sign, with the reference DFT data.
(Clearly, since the $2^\mathrm{nd}$ NN spins always remain parallel, the matrix $\mathcal{M}_2$ does not contribute to the polarization.)
This indicates that one should include at least up to 3$^\mathrm{rd}$ NN interactions for a correct description of the ME coupling in CrI$_3$, and that the 3$^\mathrm{rd}$ NN interaction is at least as important as the 1$^\mathrm{st}$ NN one. 

By comparing the two panels of Fig.~\ref{fig.CrI3_P_AFM}, one can notice that the polarization is in both cases perpendicular to the canting direction $\hat{s}$, and has a sinusoidal dependence on the canting angle $\theta$. 
In fact, as a consequence of the high symmetry of CrI$_3$, the electric dipole moment per Cr pair in the entire parameter space spanned by $(\alpha,\theta)$ is determined by a single material constant via 
\begin{equation}
    \mathbf{p} =  p_0 \sin 2\theta \, \, \hat{s} \times \hat{n},
\end{equation}
where $\hat{n}$ is the normal to the layer.
The quantity $p_0 =\frac{3}{2}( M_1^{11} + M_1^{22} + M_3^{11} + M_3^{22}) = 98\cdot10^{-5} ~\mathrm{e\AA}$ describes the strength of the coupling between an electric field and an optical magnon, and can be regarded as a magnon analogue to the Born effective charge of a phonon. 
The form of this dipole moment tells us that out-of-plane electric fields do not couple to the optical magnons, while in-plane electric fields do. 
As a consequence, an applied electric field will induce a non-zero anti-ferromagnetic (AFM) order parameter $\mathbf{L} = \frac{1}{2}(\mathbf{S}_1 - \mathbf{S}_2 ) = \sin\theta \hat{s}$, with $\mathbf{S}_1$ and $\mathbf{S}_2$ being neighboring spins. 
An in-plane electric field $\mathcal{E}$ leads to a tilt angle of $\tan 2\theta = \frac{2\mathcal{E}p_0}{J}$, where $J=11.6~\mathrm{meV/Cr}$ is the energy difference between the FM and AFM states. At small angles/electric fields, the AFM order parameter is proportional to the applied electric field, with $7~\mathrm{mV/\AA}$ resulting in an angle $\theta=10^\circ$ ($|\mathbf{L}| \approx 0.17$).

The data in Fig.~\ref{fig.CrI3_P_AFM} implies that if CrI$_3$ had N\'eel type AFM order, 
a magnetization would be induced linearly with the applied electric field. In fact it has been pointed out that MnPS$_3$, with the same crystallographic space group as CrI$_3$ but N\'eel type AFM order, is a linear magnetoelectric~\cite{PhysRevB.82.100408}. The calculations shown in Fig.~\ref{fig.CrI3_P_AFM} therefore illustrates how the model in Eq.~\ref{eq.P_Mtensor} can also be used to calculate the homogeneous linear magnetoelectric effect, in addition to the inhomogeneous ones which are the main topic of this work.

\section{Spin spirals}\label{sec.spirals}

\begin{figure}[h!]
	\centering
	\includegraphics[width=0.46\textwidth]{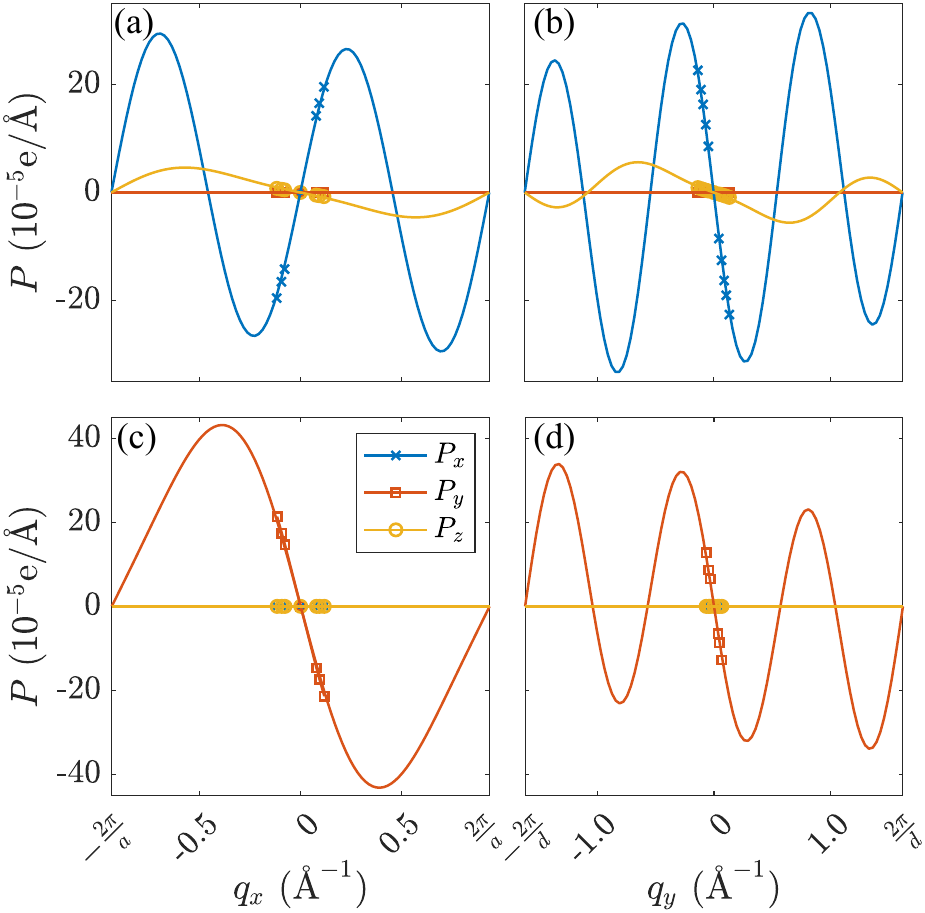} \\
	\caption{Electric polarization as a function of spin cycloid (top) or helix (bottom) wavevector, for propagation in $x$ (left) or $y$-direction (right), in a monolayer of CrI$_3$. 
 }
	\label{fig.CrI3_P_q_spirals}
\end{figure}

Next, we study the electric polarization of spin spirals.
To define the structures, we use an orthogonal triad of unit vectors $\hat{s}$, $\hat{t}$ and $\hat{n}$ in such a way that $\hat{s}={\bf q}/q=\left(\cos\alpha , \sin\alpha, 0 \right)$ is the propagation direction ({\bf q} is the wave vector of the spiral), $\hat{n}$ is the normal to the layer plane and $\hat{t}=\hat{n}\times \hat{s}$.
We focus on two types of spirals, relevant in CrI$_3$: cycloids, rotating in the $t=0$ plane, and helices, rotating in the $s=0$ plane.
The canting angle at a given magnetic site $i$ is defined in all cases by $\theta_i={\bf q}\cdot {\bf R}_{i}$.
Fig.~\ref{fig.CrI3_P_q_spirals} shows the electric polarization of cycloids (top) and helices (bottom), propagating in either $x$ ($\alpha=0$) or $y$-direction ($\alpha=\pi/2$).
Again, the results of our full model (Eq.~\ref{eq.P_Mtensor}, solid curves) are in excellent agreement with the direct DFT calculations (symbols) of the polarization for the spin cycloids and helices, which we obtained by using supercells of length $L=\frac{2\pi}{q}$.
 
At difference with the case of the optical magnon, the 2$^\mathrm{nd}$ NN interactions largely dominate the polarization of the spiral structures considered here, while the  1$^\mathrm{st}$ and 3$^\mathrm{rd}$ NN interactions have a comparatively smaller effect (except for $P_z$ of the cycloids, where $\mathcal{M}^1_{31}$ gives a larger contribution than $M^2_{32}$ in Eq.~\ref{eq.fz_M}).
This fact can be most clearly appreciated by looking at panel (b) and (d): The appearance of three maxima and three minima for the polarization on the interval $q_y \in [-\frac{2\pi}{d}, \frac{2\pi}{d} ]$ is entirely due to the longer-range 2$^\mathrm{nd}$ NN interactions, whereas the 1$^\mathrm{st}$ NN interactions alone would have led to only two maxima and two minima on the same interval. 
Thus, our results again caution against using only the 1$^\mathrm{st}$ NN interactions~\cite{PhysRevLett.107.157202,PhysRevB.88.054404} in this class of physical problems, which may lead to a qualitatively incorrect description. 

Remarkably, we find a large in-plane polarization for both helices and cycloids, while the out-of-plane components (only present in the latter) are comparatively much smaller. This outcome is in stark contrast with the predictions of the KNB spin current model~\cite{KNB}: %
According to the latter, the polarization of the cycloids should be parallel to $\hat{z}$, whereas that of the helices should be zero. 
The origin of the discrepancy is due to the assumption of cubic symmetry that is inherent to the KNB model. 
To see why in CrI$_3$ such an approach is not valid, it is useful to reason in terms of the continuum counterpart of our microscopic magnetoelectric model.
As we mentioned in Sec.~\ref{sec.CrI3}, $\tilde{f}_\mathrm{ip}=\tilde{f}_{11,31}$ is symmetry-allowed in CrI$_3$, and this component is precisely the one responsible 
 for the in-plane polarization of both helices and cycloids.
 (In fact, in the continuum limit both helices and cycloids have the same in-plane component of polarization linear in $q$, and they only deviate at larger $q$.)
The inappropriateness of KNB for noncubic crystal classes was evident already from experimental observations of electric polarization due the helical spin spirals~\cite{PhysRevB.73.220401}, motivating the development of generalized models~\cite{PhysRevLett.107.157202} such as Eq.~\ref{eq.P_Mtensor}. More recent examples include also spin cycloids with polarization components beyond the KNB model~\cite{PhysRevLett.132.086802}. The symmetry based discussion and multiscale approach taken here is useful to understand which magnetoelectric effects can be expected depending on which crystal class and spin textures are involved. For example, in Sec.~\ref{sec.Skyrmion} we will see how the $\tilde{f}_\mathrm{ip}=\tilde{f}_{11,31}$, which gives rise to the in-plane polarization of spin cycloids and helices, is precisely that which yields an in-plane polarization of anti-Skyrmions. 

The electric polarization induced by non-collinear magnetic order, such as spin-spirals, has been extensibely studied in the field of multiferroic materials, with experimentally measured polarization values ranging from $10^{-4}~\mathrm{\mu C/cm^2}$ in (Eu/Y)MnO$_3$ to $0.12~\mathrm{\mu C/cm^2}$ in DyMnO$_3$~\citep{PhysRevLett.98.057206,Tokura_spiralmulti}. 
This can be compared to the maximum values in Fig.~\ref{fig.CrI3_P_q_spirals} which correspond to $0.2~\mathrm{\mu C/cm^2}$, well above what has previously been found in spin-spiral multiferroics, and it can be attributed to the strong SOC of I. Polarization values as large as $6.1~\mathrm{\mu C/cm^2}$ have been theoretically predicted in the E-type AFM state of HoMnO$_3$~\cite{picozzi_homno3_prl2007,Picozzi_2009}, where, however, the microscopic origin of polarization has to be ascribed to Heisenberg exchange and not to the spin-orbit-driven "inverse-DMI" mechanism, as in the CrI$_3$ case.

\section{Domain Walls}\label{sec.DW}

\subsection{Structure and energetics}

The above results for helical and cycloidal spin states confirm the accuracy of the 3$^\mathrm{rd}$ NN atomistic model in describing the electric polarization of non-collinear magnetic textures in CrI$_3$, whereby the continuum model will by construction also accurately describe the polarization of these textures if the magnetization is slowly varying. Thus, we now have confidence to apply the models for calculating the electric polarization of other spin configurations, beyond reach of direct DFT calculations, such as magnetic DWs. 
An isolated magnetic DW is characterized by rotation of the magnetization from $+\hat{z}$ to $-\hat{z}$ when moving along one direction 
$\hat{s} = (\cos\alpha,\sin\alpha,0)$, parameterized by a coordinate $s=\mathbf{r}\cdot \hat{s}$, where the spins rotate in a plane normal to $\cos\beta \hat{t} - \sin\beta \hat{s}$. Then, $\beta = 0$ corresponds to a N\'eel type DW and $\beta=\pi/2$ corresponds to a Bloch type DW. 
Minimization of Eq.~\ref{eq.mag_energy} for the corresponding boundary conditions gives a spin rotation profile with angle $\theta(x) = 2\arctan(e^{x/\xi})$ between the out-of-plane direction $\hat{z}$ and the magnetization, where the DW width is determined by the length scale $\xi$ and the DW energy is $E_\mathrm{DW}  = 4\sqrt{AK} = 1.3~\mathrm{meV/\AA}$,  with numerical values given for monolayer CrI$_3$~\cite{our_PRL}. Interestingly, adding a DMI term to the magnetic energy functional does not alter the form of the DW solution, although it does alter its relative energy.

For any magnetization that is periodic along one dimension $\mathbf{M}(0) = \mathbf{M}(L)$, there is an integer winding number 
\begin{equation}\label{eq.DWwinding}
    N = \frac{1}{2\pi}  \int_{0}^L d \theta 
\end{equation}
where $\theta$ is the angle between the magnetization and it's initial direction. %
A winding number $N$ corresponds to a sequence of $2N$ DWs, but will also be identical for a spin spiral with wave number $q=\frac{2\pi}{L}$. In the coming section we show that this winding number, together with the strength of the magnetoelectric coupling parameter, fully determines the net electric dipole moment per length. 

\subsection{Electric polarization}\label{sec.DWpol}

\begin{figure}[h!]
	\centering
	\includegraphics[width=0.5\textwidth]{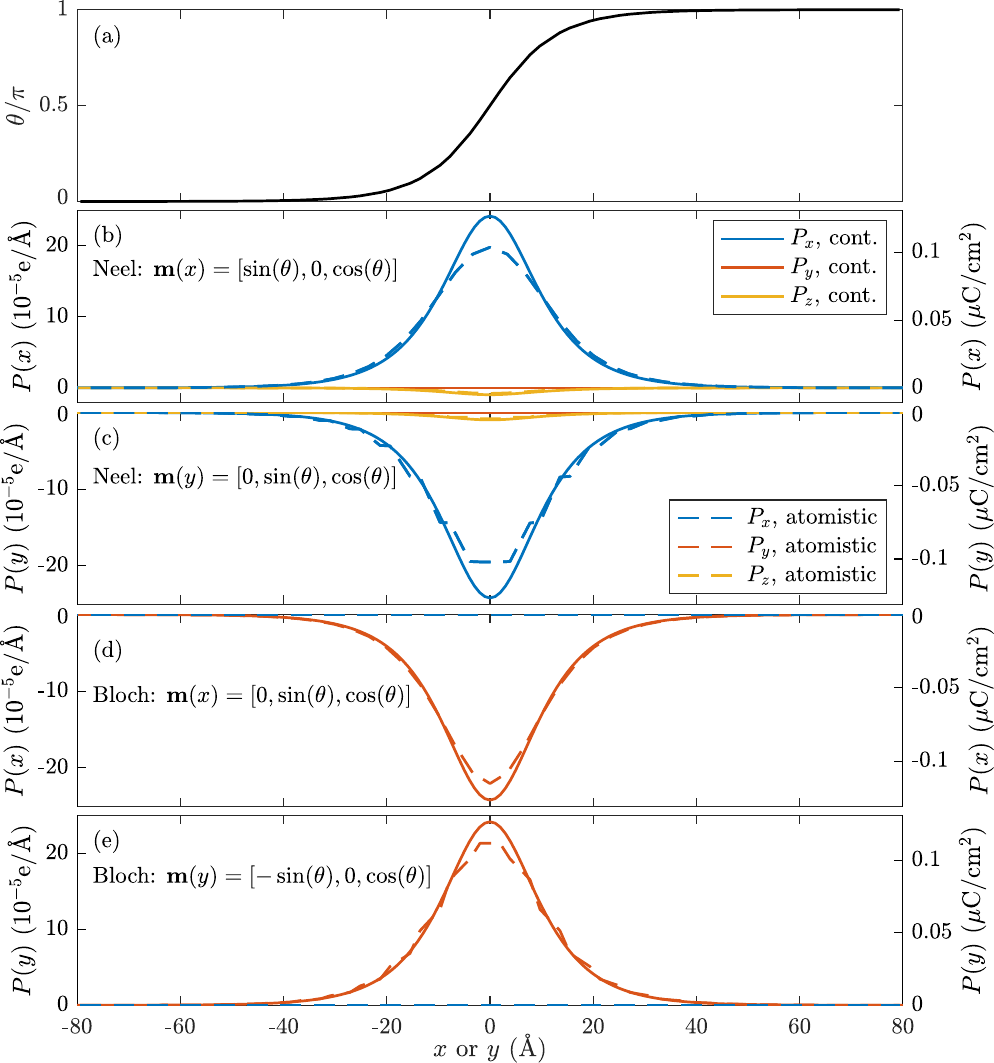} 
	\caption{(a) Spin rotation profile $\theta (s) = \arctan(\ee^{s/\xi})$ of a DW in CrI$_3$, with $\xi=8~\text{\AA}$. (b)-(d) Electric polarization as a function of position along Bloch or N\'{e}el DWs, oriented along either $x$ or $y$-direction, evaluated with either continuum (solied lines) or atomistic (dashed linens) model. 
 }
	\label{fig.CrI3_P_DW}
\end{figure}

From Eq.~\ref{eq.contmod} a DW according to the description above has a polarization
\begin{align}
\mathbf{P}(s) = \theta^\prime (s) \left[ \tilde{f}_\mathrm{ip}\cos(3\alpha+\beta) \hat{s} \right. & -  \tilde{f}_\mathrm{ip}\sin(3\alpha+\beta) \hat{t} \nonumber \\ \left. + \tilde{f}_z \cos\beta \hat{n} \right]  \label{eq.P_DWs}
\end{align}
with a coordinate system such as that introduced in the previous Sec.~\ref{sec.spirals}. Given the rotational profile $\theta(s) = 2\arctan(e^{s/\xi})$ introduced above, $\theta^\prime (s) = {\mathrm{sech}(s/\xi)}/{\xi}$. 
The in-plane component of Eq.~\ref{eq.P_DWs} clearly embodies the three-fold rotational symmetry of the CrI$_3$ crystal, in resemblance to the expression for flexoelectric polarization in Ref.~[\onlinecite{springolo2023inplane}].
The solid curves in Fig.~\ref{fig.CrI3_P_DW}(b)-(e) show the resulting electric polarization profile in monolayer CrI$_3$, 
either for $\alpha = 0$ ($s=x$) or $\alpha = \pi/2$ ($s=y$), and for N\'eel or Bloch type DWs. The magnitude and direction of the polarization for the N\'eel or Bloch walls are similar to the cycloid and helix cases in Fig.~\ref{fig.CrI3_P_q_spirals}, respectively, since the rotational sense of the spins is similar. Accordingly, the N\'eel wall polarization has the largest component along the $\hat{x}$-direction, with a small $P_z$, while the Bloch walls have $\mathbf{P} \parallel \hat{y}$. In both cases, the in-plane polarization changes sign if the DW is rotated by 90$^\circ$. The maximum magnitude is also similar for the different cases, just below $10^{-5}\mathrm{e/\AA}$. Converted into volume polarization using the I layer distance as thickness, this corresponds to $0.1~\mu\mathrm{C/cm}^2$.

As a further test of our methodology, in all cases we recalculated the polarization profile by using Eq.~\ref{eq.P_Mtensor} directly (we assume that the on-site values of the spins match the value of the continuum solution at the corresponding location); the results are plotted in Fig.~\ref{fig.CrI3_P_DW}(b)-(e) as dashed lines.
While results agree reasonably well, there is a non-negligible discrepancy at the center of the DW, where the magnetization is varying most rapidly. This can be expected for a relatively thin DW, as the  continuum model is only accurate for slowly varying magnetizations. We have confirmed (not shown) that by increasing $\xi$ the agreement between the atomistic and continuum models becomes excellent, as it should for magnetizations varying slowly relative to the interatomic distance.

As previously pointed out by Mostovoy~\cite{PhysRevLett.96.067601}, another implication of the continuum model in Eq.~\ref{eq.contmod} is that the net electric dipole moment over a DW depends on the rotation angle while it is independent of DW profile and width.  This can be anticipated from Eq.~\ref{eq.P_DWs}, where the $s$-dependence of the form $\theta^\prime (s)$ is trivially integrated to $\int_{-\infty}^\infty \theta^\prime (s) \dd s = \pi$, so for e.g. $\beta = 0$ the dipole moment per unit length is $\pi \left[ \tilde{f}_\mathrm{ip}\cos(3\alpha) \hat{s} -  \tilde{f}_\mathrm{ip}\sin(3\alpha) \hat{t} + \tilde{f}_z \hat{n} \right]$. This can be generalized to an arbitrary periodic spin texture with  winding number given by Eq.~\ref{eq.DWwinding}.
The net dipole moment per unit length of such a magnetization texture is proportional to $N$, depending only on the winding (and magnetoelectric parameters), not on the form of the function $\theta(x)$. Thus, the dipole moment of a DW is exactly half that of a spin spiral over one period. This holds in the continuum limit, where the spiral has a polarization which is linear in $q$, so that the polarization multiplied by the wavelength (the dipole moment per length) remains constant. 

\subsection{Discussion}

DWs  bring in interesting new considerations in 2D monolayer magnets. While domain formation is ubiquitous in 3D bulk magnets, 2D monolayers with perpendicular anisotropy are persistently observed to form monodomains in experiments~\cite{doi:10.1063/5.0062541,Wahab_CrI3_DWs,Fujita2022,Sun2021}. For example, in Fe$_3$GaTe$_2$ films, domain structure is observed down to thicknesses of 17.8 nm~\cite{10.1063/5.0159994}, below which a monodomain forms. This is consistent with simple arguments regarding magnetostatic dipole-dipole interactions in 2D vs 3D as well as established theoretical results for domain formation in thin films with perpendicular magnetization; there exists a critical thickness under which magnetic films with perpendicular magnetic anisotropy only form monodomains~\cite{PhysRev.70.965,Hubert_DW,Knupfer2019}. For materials with strong perpendicular anisotropy, which is typically considered an important ingredient to overcome the Mermin-Wagner theorem and stabilize magnetic order in 2D magnets, a monodomain state is expected if the thickness is much smaller than $l_c = \frac{E_\mathrm{DW}}{2K_d}$, where $E_\mathrm{DW}$ is the DW energy and $K_d=\frac{1}{2}\mu_0 M_\mathrm{s}^2$ the magnetic depolarizing field energy density for saturation magnetization $M_\mathrm{s}$~\cite{Hubert_DW}. 

To see the reason, it is useful to observe that $l_c$ coincides, modulo a dimensionless prefactor of the order of unity, with the Kittel's length. In other words, Kittel's law would predict a domain periodicity of the order of $w=\sqrt{l_c d}$, where $d$ is the thickness of the sample.
If $l_c \gg d$, one obtains a value of $w$ that is much larger than the film thickness, and the range of validity of Kittel's law breaks down. 
Indeed, in such a regime the stray fields that are responsible for the magnetostatic energy cost are no longer confined to the surface, but affect the whole volume of the film.
For the same reason, spaced-apart DW's are hardly effective at reducing the depolarizing energy within the domains, and there is no driving force for their formation.

Considering material parameters for CrI$_3$~\cite{our_PRL}, assuming an interlayer distance of $D=7~\mathrm{\AA}$~\cite{Huang2017}, one finds $l_c \approx 700D$ and a monolayer is clearly expected to form a single domain state. In practice, domains have been observed in samples as thin as $25~\mathrm{nm} = 36D$~\cite{Wahab_CrI3_DWs}. 
Given that the DW energy is $E_\mathrm{DW}=4\sqrt{AK}$, where $A$ is the spin stiffness and $K$ the magnetic anisotropy, and that a sizeable $K$ is typically considered needed to stabilize magnetism in the 2D limit, domain formation in atomically thin samples would only be expected for extremely small values of $A$, i.e. if magnetism is entirely stabilized by anisotropy, with negligible exchange interactions. 
This raises the question of whether applied electric fields can be used to stabilize magnetic DWs in 2D monolayer magnets such as CrI$_3$.
Once stabilized, the DWs should remain due to their topological protection, and could then be manipulated and/or detected by electrical means. 

The results presented in Sec.~\ref{sec.DWpol} allow us to provide a quantitative answer to this question. %
Since, as shown above, the dipole moment is independent of DW width and rotation profile, an electric field is not expected to affect the shape and size of the DW, assuming that the field is homogeneous over the region where the DW is located. 
Given the total magnitude of the dipole moment per unit length ($\pi \tilde{f}_\mathrm{ip}$ or $\pi \tilde{f}_z$ for the in-plane and out-of-plane components, respectively) and DW energy $E_\mathrm{DW}$, an estimate of the electric field needed to stabilize domains is $\mathcal{E} = \dfrac{E_\mathrm{DW}}{\pi \tilde{f}_\mathrm{ip}} = 0.20~\mathrm{V/\AA} = 2\cdot 10^4~\mathrm{kV/cm}$ for an in-plane electric field or $\mathcal{E} = \dfrac{E_\mathrm{DW}}{\pi \tilde{f}_z} = 5.0~\mathrm{V/\AA} = 5.0\cdot 10^5~\mathrm{kV/cm}$ for an out-of-plane field. 
While these are large values of electric fields, taking into consideration magnetostatic dipole-dipole interactions would favor domain formation and lower the electric fields required. 
Moreover, even an applied field insufficient to make domains energetically favored will increase the probability of their formation during a cooling process. In atomistic spin dynamics simulations, domains formed in CrI$_3$ monolayers during rapid, zero field cooling, from above $T_C$ to 0 K in 2 ns.~\cite{Wahab_CrI3_DWs} A combination of an applied electric field and rapid cooling should therefore be a viable option to experimentally induce magnetic domains in 2D magnets. If the electric fields can be applied locally, this will also allow control of the position of the DWs.  
Our calculated electric field induced DMI parameters provide the necessary prerequisites for future further investigations in this direction using atomistic spin dynamics simulations.

In thicker CrI$_3$ crystals, of a couple of nanometers or thicker, magnetic DWs do form experimentally and the layers couple ferromagnetically.
Our result imply that 
an electric polarization should appear at these magnetic DWs as well, in principle allowing for their detection and manipulation %
by electrical means. The precise form of the polarization in thicker samples may, however, differ from that of the monolayers studied here, because of inter-layer coupling and stacking dependent symmetry breaking.

A few earlier works~\cite{PhysRevLett.125.067602,PhysRevLett.125.067602,C9TC02501D,doi:10.1063/1.4917560} have presented computational predictions of the electric polarization of magnetic DWs. The predicted values are mostly of a similar order of magnitude as that found here, ranging from $0.07~\mu\mathrm{C/cm}^2$~\cite{PhysRevLett.125.067602} to $0.7~\mu\mathrm{C/cm}^2$~\cite{C9TC02501D,doi:10.1063/1.4917560}. However, they all rely on a non-relativistic (i.e. neglecting SOC) magnetostrictive phenomena due to the symmetry breaking of a collinear magnetic order. 
In most magnetic materials this is not a realistic description of a magnetic DW, where the magnetization gradually rotates over distances much larger than the interatomic spacing. As such, the predictions made here provide a more realistic description of the electric polarization of magnetic DWs than previously available. Unfortunately, little experiment data for magnetic DW polarization is available so far. \\ 

\section{Skyrmions}\label{sec.Skyrmion}

Magnetic Skyrmions are another type of magnetic topological soliton where the magnetization texture breaks inversion symmetry. 
This leads to an electric polarization in insulating materials, as has been experimentally observed in GaV$_4$S$_8$~\cite{doi:10.1126/sciadv.1500916}. In atomically thin magnets, the out-of-plane polarization may survive even in metallic materials. The electric polarization of Skyrmions in GaV$_4$S$_8$ was explained by atomistic magnetoelectric models similar to Eq.~\ref{eq.P_Mtensor}, but with additional symmetric contributions appearing due to the broken SI symmetry~\cite{doi:10.1126/sciadv.1500916,PhysRevB.99.100401} (see also App.~\ref{sec:app_macroP}). 
As we will see in the following, the continuum magnetoelectric model is particularly insightful in describing the electric polarization and the net electric dipole moment carried by Skyrmions.

\subsection{Magnetization texture}

\begin{figure} 
	\centering
	\includegraphics[width=0.49\textwidth]{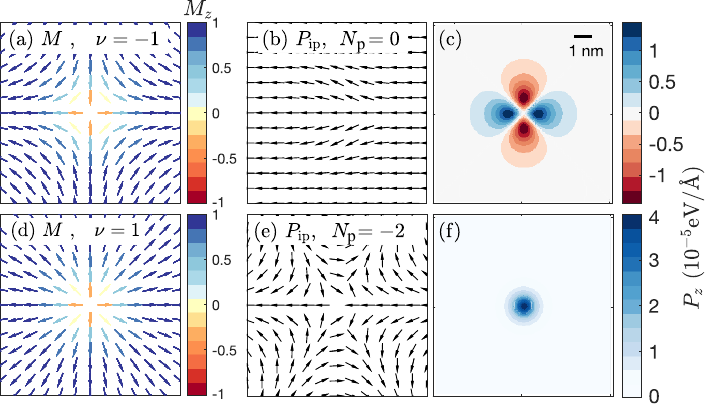} \hspace{0.02\textwidth}%
	\caption{The Magnetization texture and electric polarization for an  anti-Skyrmions ($\nu=-1$) in the top row and a Skyrmion ($\nu=1$) in the bottom row, both with $\gamma=0$. The first column shows the magnetization with the in-plane direction as arrows and the out-of-plane component represented by color. The middle column shows the in-plane direction of the electric polarization, and the third column the out-of-plane component of polarization, $P_z$. 
 }
	\label{fig.Skyrm_pol}
\end{figure}

A Skyrmion magnetization texture is conveniently described in polar coordinates $\mathbf{M}(\mathbf{r}) = \left( \sin\Theta \cos\Phi , \sin\Theta \sin\Phi , \mathcal{P} \cos\Theta \right)$, 
at position $\mathbf{r} = (r \cos \phi , r \sin \phi)$, relative to the center of the Skyrmion~\cite{GOBEL20211,Bogdanov2020}.
The polar angle $\Theta(r,\varphi)$ fulfills $\Theta(0,\varphi)=\pi$ and $\Theta(r,\varphi) \rightarrow 0$ for large $r$. The azimuthal angle $\Phi(r,\varphi)$ is periodic $\Phi(r,\varphi) = \Phi(r,\varphi + 2\pi \nu)$ with integer $\nu$. We  consider Skyrmions with $\Theta=\Theta(r)$ only dependent on $r$, and $\Phi(\varphi) = \nu \varphi + \gamma$, as is appropriate for a continuum description of an isolated Skyrmion in a material that is isotropic in the plane, such as CrI$_3$. Here, $\nu$ is vorticity (integer), and $\gamma$ is helicity, while $\mathcal{P} =\pm 1$ is polarity. 
Skyrmions are characterized by the integer Skyrmion number 
\begin{equation}
    Q = \frac{1}{4\pi} \int d^2\mathbf{r}~ \mathbf{M} \cdot \left( \frac{\partial \mathbf{M}}{\partial x} \times \frac{\partial \mathbf{M}}{\partial y} \right) = \nu \mathcal{P} .
\end{equation}
We will primarily consider $\mathcal{P}=+1$ and denote $Q=\nu=+1$ structures as Skyrmions, and those with $Q=\nu=-1$ as anti-Skyrmions. 

In order to illustrate a number of basic properties of the (anti)-Skyrmions, in the next subsections we shall initially assume a fixed radial form factor of the skyrmion that is independent of its size. 
This means using the following \emph{Ansatz} for $\Theta(r)$,
\begin{equation}
\label{ansatz1}
    \Theta(r) = \theta(r/w),
\end{equation}
where $w$ defines the length scale.
This assumption yields a qualitatively correct picture for most properties we are interested in, and allows for an easier understanding of the main mechanisms at work. 
In practice, we shall use the form~\cite{Bogdanov2020} 
\begin{equation}
\label{ansatz2}
    \theta(x) = 4\arctan(\mathrm{e}^{-x}),
\end{equation} 
which mimicks the DW solution derived in the previous section. The rationale behind this choice lies in the fact that the skyrmion, in the limit of a large size, resembles a DW structure with cylindrical symmetry.
Note that, unless otherwise specified, our formal results in the remainder of this Section do not assume either Eq.~\ref{ansatz1} or Eq.~\ref{ansatz2}.
In any case, we shall relax this assumption in Sec.~\ref{sec.varskyrmsol}, where we benchmark our conclusions against the exact solution for an isolated Skyrmion emerging from the continuum equations.

A representative illustration of the magnetization textures associated to an anti-Skyrmion ($\nu=-1$) and a Skyrmion ($\nu=+1$) are shown in Fig.~\ref{fig.Skyrm_pol}(a) and (d), respectively, with arrows showing the in-plane component of magnetization and colors showing the out-of-plane component. The case shown is $\gamma=0$, i.e. a N\'eel Skyrmion, whereas changing the value to $\gamma=\pi/2$ would rotate the spins to form a Bloch Skyrmion.

\subsection{Polarization texture}

Application of Eq.~\ref{eq.contmod} to the Skyrmion magnetization, in the appropriate polar coordinates, leads to an electric polarization
\begin{widetext}
\begin{align} \label{eq.P_skyrm}
    \mathbf{P}  = & \mathcal{P}  \left( - \tilde{f}_\mathrm{ip} \cos\left[ (\nu+1)\varphi +\gamma \right] F_\nu^-(r) ~ , ~ \tilde{f}_\mathrm{ip} \sin\left[ (\nu+1)\varphi +\gamma \right] F_\nu^-(r) ~ , ~ \tilde{f}_z \cos\left[ (\nu-1)\varphi +\gamma \right] F_\nu^+(r) \right) \nonumber \\  & + \tilde{f}_{12,12}\frac{\nu}{r}\sin^2\Theta \left( \cos\varphi , \sin \varphi, 0 \right), ~ ~ \text{where} ~ ~ ~
      F_\nu^\pm (r)   = \frac{1}{2r} \left\{ \nu \sin{(2\Theta)} \pm 2 r \Theta'(r) \right\}, %
\end{align}
\end{widetext}
from which several interesting observations can be made. 
First, we can characterize the topological properties of the in-plane components of ${\bf P}$ via the concept of vorticity~\cite{RevModPhys.95.025001}
\begin{equation}\label{eq.windingN}
    N_P = \frac{1}{2\pi} \oint_C d\mathbf{l} \cdot \frac{d \Phi_P}{d\mathbf{l}} = -(\nu+1)
\end{equation}
 where $\Phi_P$ is the angle between the polarization and $\hat{x}$, and the closed path $C$ is arbitrary provided that it contains the center of the Skyrmion. 
 (This is in all respects equivalent to Eq.~\ref{eq.DWwinding}, only applied to a closed path in open boundary conditions.)
 Thus, the (in-plane projection of the) polarization, which has contributions from two terms, respectively due to $\tilde{f}_\mathrm{ip}$ and $\tilde{f}_{12,12}$, inherits a $\nu$-dependent vorticity from the topology of the magnetization texture. 
 The first term alone would give precisely $N_P=-(\nu+1)$, whereas the second term alone would give a $\nu$-independent winding of $N_P=1$, which however vanishes when both terms are considered together. That is, taking the integral in Eq.~\ref{eq.windingN}, including both terms, along, for example, a circular path at small $r$, one finds the result $N_P = -(\nu+1)$ determined entirely by the winding of the dominating first term.

The right part of Fig.~\ref{fig.Skyrm_pol} shows the polarizations from Eq.~\ref{eq.P_skyrm} (right), for an anti-Skyrmion with $\nu=-1$ (top) and a Skyrmion with $\nu=1$ (bottom). Fig.~\ref{fig.Skyrm_pol}(b) and (e) illustrate the $N_P=0$ and $N_P=-2$ polarization vorticity of the Skyrmion and anti-Skyrmion, respectively. Integration of the in-plane polarization yields a non-zero net dipole moment only when $N_P=0$, i.e. for the anti-Skyrmion with $\nu=-1$, while it is zero for the Skyrmion. Thus, we expect that the net electric dipole moment (areal integral of the polarization) should point in-plane for the anti-Skyrmion. 
The out-of-plane polarization component $P_z$, shown in Fig.\ref{fig.Skyrm_pol}(c) and (f), instead has an angular dependence of the form $(\nu-1)\varphi$, as seen in Eq.~\ref{eq.P_skyrm}. Hence,
$P_z$ of the anti-Skyrmion, shown in Fig.~\ref{fig.Skyrm_pol}(c), has equal positive and negative regions, and its integral, i.e. the net dipole moment in the $z$-direction, is zero. The Skyrmion case in Fig.\ref{fig.Skyrm_pol}(f), on the other hand, has full 2D rotational symmetry and a non-zero out-of-plane dipole moment. 

\subsection{Dipole moment}

We confirm the above observation by calculating the net electric dipole moment of a general Skyrmionic magnetization texture
\begin{equation}
\mathbf{p} = \int \mathbf{P}(\mathbf{r} ) \dd^2\mathbf{r}  = \left\{
    \begin{aligned}
    &  p_\mathrm{ip}\left( -\cos\gamma, \sin\gamma, 0\right)\mathcal{P} , && \text{if}\ \nu = -1  \\
    &  p_z\left( 0,0,\cos\gamma \right)\mathcal{P} , && \text{if}\ \nu = +1 \\ 
    & 0, && \text{if}\ \nu \neq \pm1
  \end{aligned} \right.
\end{equation}
where $p_\mathrm{ip}=2\pi \int rF_{-1}^-(r)\dd r$ and $p_z=2\pi \int rF_{1}^+ (r)\dd r$ are the magnitudes of the in-plane and out-of-plane Skyrmion dipole moments (the $z$-component is also scaled by $\cos\gamma$). Thus, a non-zero net electric dipole moment only appears for Skyrmions and anti-Skyrmions with $\nu=\pm 1$, while it is zero for all higher-order (anti-)Skyrmions ($|\nu|>1$). The anti-Skyrmion ($\nu=-1$) has an in-plane dipole moment with direction determined by $\gamma$, while the Skyrmion $\nu=1$ has an out-of-plane dipole moment with magnitude and direction determined by $\gamma$. In both cases, the polarity $\mathcal{P} $, flips the sign of the electric dipole moment. These results hold qualitatively for any material with the same crystal symmetry as CrI$_3$ and any Skyrmion radial profile $\Theta(r)$, while it can differ for other crystal symmetries, where the results should be re-derived using the correct form of the tensor $\tilde{f}_{ij,\alpha\beta}$.  The high-symmetry (e.g. cubic) case is obtained by setting $\tilde{f}_{12,12}=\tilde{f}_z$ and $\tilde{f}_\mathrm{ip}=0$. Then, only the net out-of-plane electric dipole moment for Skyrmions ($\nu=+1$) remains non-zero, which is a universal feature of all (insulating or 2D) materials hosting Skyrmions. The in-plane dipole moment of anti-Skyrmions, on the other hand, vanishes in the cubic case and is non-zero in specific crystallographic point groups where $\tilde{f}_\mathrm{ip}$ is symmetry allowed (See Appendix~\ref{app.contmodel}).

In the specifics of our fixed-shape skyrmion model, the dipole moment is a linear function $p_i=d_iw$ of the size $w$, where $d_i$ is proportional to $f_i$, and $i$ is either $z$ or in-plane. 
To see this, we can define a Skyrmion radius $w$ in line with Eq.~\ref{ansatz1} and write the radial profile in terms of a dimensionless radial coordinate $\rho = r/w$, $\Theta(\rho) = \Theta(r/w)$; then, derivatives yield factors $1/w$ while surface integration gives $w^2$ 
(See Eq.~\ref{eq.E_Efield_rho} for an expression for $d_i$ in terms of $\Theta(r)$ and the materials parameters).
Using the ansatz in Eqs~\ref{ansatz2}, we find numerical values $d_\mathrm{ip} = 37\cdot 10^{-3}e$ and $d_z = 1.5\cdot 10^{-3}e$ for the in-plane dipole moment of the anti-Skyrmion or out-of-plane dipole moment of the Skyrmion, respectively. 
The results imply that large (anti-)Skyrmions carry a large electric dipole moment. We shall see in the following Section that this leads to a large amount of energy by coupling to an applied electric field.

\subsection{Energetics}

\begin{figure} 
	\centering
	\includegraphics[width=0.45\textwidth]{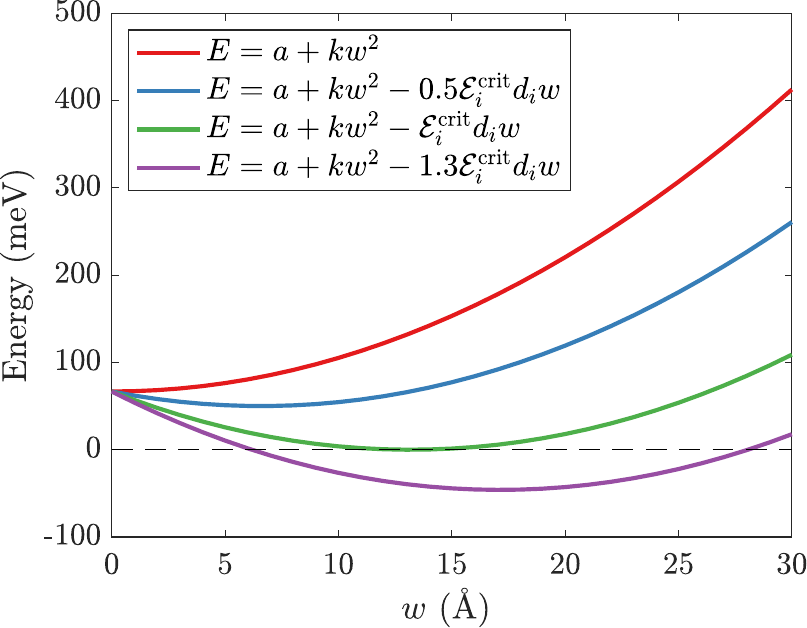} 
	\caption{Energy as a function of Skyrmion size $w$, relative to the ferromagnetic state, evaluated for a (anti-)Skyrmion with $\Theta(r) = 4\arctan(\mathrm{e}^{-r/w})$, considering exchange and anisotropy energy, and coupling to an applied electric field, expressed in terms of the critical field $\mathcal{E}_i^\mathrm{crit} = 2\sqrt{a k} / d_i$.  
 }
	\label{fig.Skyrm_p_w}
\end{figure}

In absence of DMI/electric fields, the magnetic energy (Eq.~\ref{eq.mag_energy}) of an (anti-)Skyrmion in a material with uniaxial magnetic anisotropy such as CrI$_3$ is, within the length-scale {\em Ansatz} of Eq.~\eqref{ansatz1},
\begin{align}\label{eq.E_skyrm_nofield}
E & = E_\mathrm{Sk} - E_\mathrm{FM} =  a + k w^2  , \quad a,~ k>0.  
\end{align}
This follows from the same arguments as those above, which led to the dipole moment being proportional to $w$. 
Within the specific choice of form factor given by Eq.~\eqref{ansatz2}, we find numerical values of $a = 0.4~\mathrm{meV}$ and $k=0.4~\mathrm{meV/\AA^2}$ (see Appendix~\ref{app.Skyrm_sol}).
Eq.~\ref{eq.E_skyrm_nofield} implies that if a (anti-)Skyrmion were to form, it would shrink to zero radius (i.e., the lowest-energy state respecting the topology). 
A numerical illustration of this outcome is provided in Fig.~\ref{fig.Skyrm_p_w}(a), where Eq.~\ref{eq.E_skyrm_nofield} is plotted as red line, with the values for $a$ and $k$ given above.
The zero radius "needle"-like (anti-)Skyrmion would have energy $a$, related to the spin stiffness.
Next, we will see how the magnetoelectric coupling of an applied electric field to the dipole moment described in the previous section affects this situation.  

As seen above, the (anti-)Skyrmions will carry an electric dipole moment $p=dw$, whereby a uniform applied electric field lowers the energy of the (anti-)Skyrmion by the dipole energy 
 \begin{equation}\label{eq.edip_energy}
     E_\mathrm{dip} = -\mathcal{E} \int \mathbf{P}(\mathbf{r} ) \dd^2\mathbf{r} = -\mathcal{E} dw, 
 \end{equation}
where  the last equality holds assuming the Skyrmion has one well defined length scale of $w$ (Eq.~\ref{ansatz1}).  As we have seen, an equivalent way to view this energy gain is that an electric field induces DMI which favors (anti-)Skyrmion formation. There will, however, be a competition between this energy gain and the other terms in the magnetic energy [magnetic exchange and anisotropy of Eq.~\ref{eq.E_skyrm_nofield}, shown with red line in Fig.~\ref{fig.Skyrm_p_w}(a)], all of which should be considered together to assess the possibility of electric field stabilization of Skyrmions.
Combining Eq.~\ref{eq.edip_energy} with the energy in Eq.~\ref{eq.E_skyrm_nofield}, leads to a simple second order polynomial $E(w) = a - \mathcal{E}_i d_i w + k w^2$ for the energy as a function of Skyrmion size $w$ under applied electric field $\mathcal{E}_i$, in direction $i$. 
This is also plotted for a few electric field values in Fig.~\ref{fig.Skyrm_p_w}, from which we can now present simple schematic arguments regarding the stability of (anti-)Skyrmions in electric fields; at zero field, the lowest energy Skyrmion state is that with zero radius. An applied electric field shifts the energy minimum to larger values of $w=\frac{d}{2k}\mathcal{E}$, leading to metastable Skyrmion states with radius proportional to the magnitude of the field. At sufficiently large electric fields, the energy minimum becomes negative, and (anti-)Skyrmions are lower in energy than the FM state. 
The critical field in this model is $\mathcal{E}_i^\mathrm{crit} = 2\sqrt{a k} / d_i$, where the energy minimum is exactly at zero, so a single Skyrmion state is degenerate with the FM state. Numerically, with the materials parameters found for CrI$_3$, we find $\mathcal{E}_\mathrm{ip}^\mathrm{crit} = \frac{2\sqrt{ak}}{d_\mathrm{ip}} = 0.27~\mathrm{V/\AA}$ for an in-plane field stabilizing an anti-Skyrmion, and $\mathcal{E}_z^\mathrm{crit} = \frac{2\sqrt{ak}}{d_z} = 6.7~\mathrm{V/\AA}$ for an out-of-plane field stabilizing a Skyrmion. 
Note that in both cases the Skyrmion size at the critical field is $w = \sqrt{a/k} \approx 1.646 \sqrt{A/K} = 13~\mathrm{\AA}$.
While these are very large electric fields, at least the one for the in-plane case may be realistically achievable in experiments. Smaller electric fields, would already be sufficient to stabilize finite size Skyrmions as metastable states. 
Moreover, these critical fields are expected to be overestimated because of magnetostatic dipole interactions which would lower the energy of these spin textures relative to the ferromagnetic state. Additionally, the calculations relied on the radial profile $\Theta(r) = 4\arctan(\mathrm{e}^{-r/w})$ which does not describe the true stationary state of the (anti-)Skyrmion solutions, therefore overestimating their energies. In the following Section, we improve our description by numerically minimizing the magnetic energy functional under applied electric field, providing a correct radial profile.

\subsection{Variational Skyrmion solution}\label{sec.varskyrmsol}

\begin{figure} 
	\centering
	\includegraphics[width=0.48\textwidth]{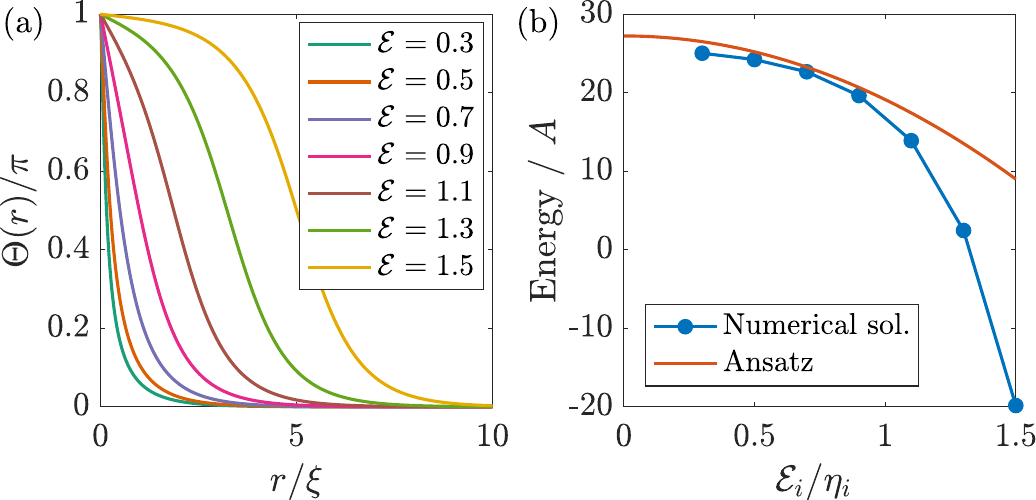} \\
	\caption{(a) The radial profile $\Theta(r)$ that solves Eq.~\ref{eq.EL} for (anti-)Skyrmions at a certain electric field $\mathcal{E}$ in units of $\eta_i$, corresponding to an induced DMI $ \mathcal{D}_i = \left| \tilde{f}_{i} \right| \mathcal{E}_i$  ($i$ denoting $z$ or in-plane, depending the field direction). (b) Energy, relative to the FM state, of the solutions presented in (a) for (anti-)Skyrmions as function of applied field (blue line, circles). The result $E = a - \frac{(d\mathcal{E})^2}{4k}$ from the Ansatz $\Theta(r) = \arctan(\mathrm{e}^{-r/w})$ is shown for comparison (orange line). 
 }
	\label{fig.Skyrm_sols}
\end{figure}

The energy density contribution from the coupling of a (uniform) electric field to the magnetic texture is given by the
middle expression in Eq.~\ref{eq.edip_energy} (equivalent to an electric field induced DMI term) added to Eq.~\ref{eq.mag_energy}.
By numerically minimizing the energy functional (see Appendix~\ref{app.Skyrm_sol}) we find solutions for $\Theta(r)$, plotted in Fig.~\ref{fig.Skyrm_sols}(a), and corresponding energies in Fig.~\ref{fig.Skyrm_sols}(b). Note that the solutions only depend on the magnitude of the electric field induced DMI value $\left| \mathcal{D}_ i \right| =  \left| \tilde{f}_{i} \mathcal{E}_i \right| $. Thus, an in-plane electric field $\mathcal{E}_\mathrm{ip}$ gives an anti-Skyrmion with identical radial profile and energy as that of a Skyrmion in an out-of-plane field $\left| \mathcal{E}_z \right| = \left| \mathcal{E}_\mathrm{ip} \frac{\tilde{f}_\mathrm{ip}}{\tilde{f}_z}\right|$. With the field expressed in terms of the electric field unit $\eta_i = \frac{A}{|f_i| \xi}$, the energy plotted in Fig.~\ref{fig.Skyrm_sols}(b) is thus identical for the two cases, but the unit is different for the different field directions, $\eta_z = 3.9~\mathrm{V/\AA}$ and $\eta_\mathrm{ip} = 0.16~~\mathrm{V/\AA}$. 
The results qualitatively confirm the arguments above; the Skyrmion radius increases with the magnitude of the applied electric field, and the energy of the (anti-)Skyrmion solution decreases with increasing field, eventually allowing solutions that are lower in energy than the FM state. 
Since the electric field here is a parameter which linearly scales the magnitude of the DMI, the results can more generally be understood as that a larger DMI favors (anti)-Skyrmion formation but also increases the equilibrium size of the (anti)-Skyrmion. 
As it should, the energy of the correct numerical solution is lower than that obtained with the ansatz (Eq.~\ref{ansatz2}), which is also plotted for comparison in Fig.~\ref{fig.Skyrm_sols}(b). Hence, the critical field of the transition is also reduced in comparison to the previous estimate, to a value of $\mathcal{E}_i = 1.32 \eta_i$, or approximately $5.2~\mathrm{V/\AA}$ and $0.2~\mathrm{V/\AA}$ for the Skyrmion or anti-Skyrmion stabilized by out-of-plane and in-plane fields, respectively. 

Previous work, using DFT calculations under finite electric field in combination with atomistic spin dynamics simulations, argued that an out-of-plane electric field of $0.2~\mathrm{V/\AA}$ is sufficient to stabilize Skyrmions in CrI$_3$~\cite{Behera_CrI3_Efield}. This was questioned in subsequent work of Ghosh et al., finding that a 45 times larger field would be needed~\cite{Ghosh_CrI3_Efield}. Our results are consistent with those of Ghosh et al. in the case of an out-of-plane electric field. Interestingly, our results indicate the previously overlooked possibility, {\em i.e.}  to stabilize anti-Skyrmions using an in-plane electric field, is possible at significantly lower field strengths.

\subsection{Discussion of Energetics of the Different Phases}

For each of the non-collinear spin textures discussed above (spirals, DWs, Skyrmions) there is a critical DMI, and corresponding critical electric field, at which these structures become lower in energy relative to the FM state, as summarized in Table~\ref{table.critfield}. Interestingly, the critical field is very similar for the three types of spin states. The comparison in energy between the three types of spin states above the critical field is not straight forward as it depends on the density of DWs and Skyrmions. Spin spirals and DWs have equivalent magnetization windings and are indistinguishable at high DW densities, which would be favored at large homogeneous fields. Thus, a homogeneous electric field would likely favor a spin spiral state, rather than DWs. However, one can envisage that an electric field applied locally along a line could stimulate the appearance of an isolated DW. For the (anti-)Skyrmions, it is again difficult to give precise statements of their stability relative to spin spirals. However, Skyrmion states are routinely stabilized over spiral states using applied magnetic fields~\cite{Tokura2021}. Thus, a combination of applied electric and magnetic fields is a viable path to (anti-)Skyrmion stabilization. Alternatively, one could again consider a locally applied electric field, e.g. over a circular region, as a possible path to stabilize (anti-)Skyrmions. Further investigations in these directions is of interest for future follow-up studies.  Note that, as was shown in Fig.~\ref{fig.CrI3_P_AFM}, even a small in-plane electric field will induce a small antiferromagnetic order parameter, proportional to the field strength.
\begin{table}[]
\begin{tabular}{lcccc}
\hline
                                  & Spiral &   DW   & Skyrmion Ansatz & Skyrmion \\ \hline
$\mathcal{E}_\mathrm{ip}$ (V/\AA) &  0.22  &   0.20  &    0.27     &    0.21    \\
$\mathcal{E}_z$ (V/\AA)           &  5.5   &   5.0   &    6.7      &    5.2      \\ 
$\mathcal{D}$ (eV/\AA)            &  0.43  &  0.42   &    0.55     &    0.41     \\ \hline
\end{tabular}
\caption{Critical electric field in the in-plane or out-of-plane ($z$) direction, to stabilize spin spirals, DWs, or Skyrmions, relative to the FM state. The Skyrmion values are given based on the ansatz $\Theta(r) = 4\arctan(\mathrm{e}^{-r/w})$, and the numerical solution for $\Theta(r)$ (rightmost column). The corresponding critical DMI $\mathcal{D} = \tilde{f}_\mathrm{ip} \mathcal{E}_\mathrm{ip} = \tilde{f}_{z} \mathcal{E}_z$ is also listed. }\label{table.critfield}
\end{table}

\section{Conclusions}
We have presented a multiscale approach to magnetoelectricity, connecting atomistic and continuum descriptions, suitable at different length scales. We determined all parameters of the models from first principles DFT calculations and validated that they accurately describe the electric polarization of complex spin textures, such as spin spirals. We also showed that the parameters of the model are equivalent to the electric field induced DMI, known to stabilize a variety of non-collinear magnetic phases, including topologically protected solitons, such as Skyrmions.  Using this approach, we have studied the coupling of electric fields to topological solitons in the 2D ferromagnet CrI$_3$. In particular, we investigated the possibility to stabilize domain walls or (anti-)Skyrmions. Earlier studies regarding the effect of electric fields on 2D magnets often focused on out-of-plane fields~\cite{}. Interestingly, here we find that in-plane electric fields couple more strongly (by more than an order of magnitude) to the magnetism of CrI$_3$ than out-of-plane fields. Moreover, we showed that the net electric dipole moment of Skyrmions is out-of-plane while that of anti-Skyrmions is in-plane, whereby stabilization of anti-Skyrmions using an in-plane electric field may be a viable possibility in CrI$_3$. 

While the electric field needed to stabilize topological magnetic solitons found here are large, their formation could be stimulated by cooling samples under an applied field, after which they would be topologically protected and expected to remain in the sample. Locally applied fields then would enable the stabilization and control of Skyrmions and DWs. Moreover, in the calculations presented here, magnetostatic dipole-dipole interactions were neglected. Although these are expected to be of less importance in 2D magnets, compared to the 3D case, they could contribute to stabilizing the non-collinear spin textures considered here over the FM state. The effect of temperature and applied magnetic fields would also be interesting to study. Given that the electric field induced DMIs are available from the magnetoelectric models considered here, these topics would be feasible for future studies, e.g. using atomstic spin dynamics simulations~\cite{Eriksson2017}. In addition to their importance for electric field engineering of topological magnetic solitons in 2D magntets, the results, together with recent studies effects of curvature on magnetism~\cite{our_PRL}, provide a promising starting point for exploring the intriguing relationship between magnetoelectric effects and flexomagnetism, potentially opening up an exciting new rich research landscape with a flexomagnetoelectric interplay between magnetism, electric polarization and curvature.

\begin{acknowledgments}
A.E. acknowledges insightful discussions with F. N. Rybakov, and financial support from the Swedish Research Council (VR - 2018-06807 and 2022-04720), \AA Forsk (22-441), the G\"oran Gustafsson Foundation, Wallenberg Initiative Materials Science for Sustainability (WISE) and the Swedish e-Science Research Center. 
M.S. acknowledges the support of the State Investigation Agency through the Severo Ochoa Programme for Centres of Excellence in R\&D (CEX2023-001263-S), of the Ministry of Science, Innovation and Universities (Grant No. PID2023-152710NB-I00) and of Generalitat de Catalunya (Grant No. 2021 SGR 01519).
This work has been funded by the European Union - NextGenerationEU, under the Italian Ministry of University and Research (MUR) PRIN-2022 project  "SORBET: Spin-ORBit Effetcs in Two-dimensional magnets" (IT-MIUR Grant No. 2022ZY8HJY) and National Innovation Ecosystem grant ECS00000041 - VITALITY - CUP B43C22000470005.
Computational work was done on resources at PDC, Stockholm and NSC, Link\"oping via the National Academic Infrastructure for Supercomputing in Sweden (NAISS), as well as the Galileo system of Cineca through HPC-Europa3 (HPC17A3WLE), which also supported the collaborative effort by funding an international research visit.  
\end{acknowledgments}

\appendix

\section{Macroscopic polarization induced by spiral magnetism within the atomistic approach}\label{sec:app_macroP}

The starting point for developing an atomistic theory of spin-induced polarization is assuming that the macroscopic polarization can be decomposed as a sum of atomic (local) and pairwise contributions involving two spins, and then considering the polarization of a general non-collinear configuration of each spin dimer\cite{KNB, kaplan_prb2011, PhysRevLett.107.157202}. The dipole induced by each spin dimer can be then expanded as a Taylor series of spin components, where only the lowest orders are retained\cite{kaplan_prb2011, PhysRevLett.107.157202}. In the most general case, the macroscopic polarization can then be written as:
\begin{eqnarray}\label{eq:general_P}
\bm P &=&\frac{1}{V}\left\{\sum_i \bm S_i\cdot \bm{\mathcal{P}}^i\cdot\bm S_i + \frac{1}{2}\sum_{ij}\bm S_i\cdot \bm{\mathcal{P}}^{ij}\cdot\bm S_j\right\},
\end{eqnarray}
where $V$ is the (macroscopic) volume, $i$ runs over all lattice sites and $j$ over all neighbors for each given site, that is typically restricted to few nearest-neighbour shells, while the prefactor $1/2$ is introduced to count each bond contribution only once. $\bm{\mathcal{P}}^i$ and $\bm{\mathcal{P}}^{ij}$ relate the localized spins to polarization, being both rank-three polar tensors with elements of the form $\mathcal{P}_{\alpha\beta\gamma}$, where $\alpha$ is the polarization component induced by the coupling of spin components $\beta$ and $\gamma$.  The intrasite coupling tensor is symmetric under the exchange of the second and third Cartesian indices, while the intersite coupling tensor can be decomposed in a symmetric and antisymmetric part with respect to the  $\beta,\gamma$ indices:
\begin{eqnarray}\label{eq.intersiteP}
{\mathcal P}^{ij}_{\alpha\beta\gamma} &=& \frac{1}{2}\left[{\mathcal P}^{ij}_{\alpha\beta\gamma}+{\mathcal P}^{ij}_{\alpha\gamma\beta}  \right] +\frac{1}{2}\left[{\mathcal P}^{ij}_{\alpha\beta\gamma}-{\mathcal P}^{ij}_{\alpha\gamma\beta}   \right] \nonumber\\
&\equiv& ({\mathcal P}^{ij}_{\alpha\beta\gamma})^S+({\mathcal P}^{ij}_{\alpha\beta\gamma})^A.
\end{eqnarray}
The antisymmetric part can be expressed in a more compact form as $\bm S_i\cdot (\bm{\mathcal{P}}^{ij})^A\cdot\bm S_j = \bm{\mathcal{M}}^{ij} (\bm S_i\times\bm S_j)$, where the $\bm{\mathcal{M}}^{ij}$ is now a rank-two polar tensor (3$\times$3 matrix) with elements $\mathcal{M}_{\alpha\beta} = \epsilon_{\beta\gamma\delta}\mathcal{P}_{\alpha\gamma\delta}$, with $\epsilon_{\beta\gamma\delta}$ the Levi-Civita symbol and assuming summation over repeated indices. The latter then takes the form of a generalized KNB coupling, describing the electric polarization induced by the spin-current mechanism\cite{PhysRevLett.107.157202}. Microscopically, the intrasite contribution has been related to the spin-dependent $p$-$d$ hybridization mechanism, responsible, e.g., for magnetically induced ferroelectricity in the collinear antiferromagnet Ba$_2$CoGe$_2$O$_7$\cite{BCGO_prl2010,BCGO_prb2011}. The symmetric intersite contribution, instead, has been shown to account for the electronic contribution to  type-II multiferroicity in collinear magnets such as HoMnO$_3$\cite{PhysRevB.88.054404,picozzi_homno3_prl2007}, microscopically arising from a subtle interplay between magnetic and orbital ordering\cite{barone_prl2011}.
It also contributes to the electric polarization of Skyrmions in multiferroic GaV$_4$S$_8$~\cite{PhysRevB.99.100401}, where it is symmetry allowed due to the inversion asymmetric crystal structure.

We now analyse the general form of the macroscopic polarization induced by an helimagnetic configuration.
For a crystalline periodic lattice, the atomic positions are given by $\bm r_i\equiv\bm r_{nl} = \bm R_n +\bm \tau_l$, where $\bm R_n$ is the lattice vector of the $n^{th}$ cell and $\bm \tau_l$ denotes the atomic position within the unit cell, while $V=N\Omega$, $N$ being the total number of cells each with a volume $\Omega$ (area in 2D crystals). A general non-collinear spin modulation with wave vector $\bm q$ can be expressed as:
\begin{eqnarray}\label{eq:sdw}
\bm S(\bm r_{nl}) &=& s_1\bm e_1\cos(\bm q\cdot \bm r_{nl})+s_2\bm e_2\sin(\bm q\cdot \bm r_{nl})+s_3\bm e_3,\nonumber\\
\end{eqnarray}
where $\bm e_i$ are unit vectors forming an orthogonal basis. Eq. (\ref{eq:sdw}) describes a sinusoidal spin-density wave if only $s_1$ or $s_2$ is nonzero, an (elliptical) helix if only $s_3$ is zero, and a conical (elliptical) helix when all $s_i$ coefficients are nonzero. Plugging Eq. (\ref{eq:sdw}) in Eq.(\ref{eq:general_P}) and taking the thermodynamic limit $N\to\infty$, one gets:
\begin{widetext}
\begin{eqnarray}\label{eq:genP_sdw}
\bm P &=&\frac{1}{\Omega}\sum_l\Bigl\{\frac{1}{2}\left[s_1^2 \,\bm e_1\cdot \bm{\mathcal{P}}^l\cdot\bm e_1 + s_2^2 \,\bm e_2\cdot \bm{\mathcal{P}}^l\cdot\bm e_2 + 2s_3^2 \,\bm e_3\cdot \bm{\mathcal{P}}^l\cdot\bm e_3  \right]\Bigr. + \frac{1}{2}\sum_{j} s_3^2\,\bm e_3\cdot (\bm{\mathcal{P}}^{lj})^S\cdot\bm e_3  \nonumber\\
&+&\Bigl. \frac{1}{4}\sum_{j} \,\Bigl[ s_1^2\,\bm e_1\cdot (\bm{\mathcal{P}}^{lj})^S\cdot\bm e_1 +s_2^2\,\bm e_2\cdot (\bm{\mathcal{P}}^{lj})^S\cdot\bm e_2\Bigr] \cos(\bm q\cdot\bm r_j)+\frac{1}{2}\sum_{j}s_1s_2\,\bm{\mathcal{M}}^{lj}(\bm e_1\times\bm e_2)\,\sin(\bm q\cdot \bm r_j)\Bigr\}.
\end{eqnarray}
\end{widetext}
Both the intrasite and symmetric intersite contributions clearly describe the polarization induced by the collinear components of the spin configuration, whereas the contribution due to noncollinear components is fully accounted for by the antisymmetric intersite coupling tensor, describing a generalized spin-current (gKNB) mechanism, in agreement with earlier conclusions drawn for an isolated spin dimer\cite{kaplan_prb2011}.  Additionally, crystal symmetries can introduce further constraints on the allowed contributions to macroscopic polarization. The symmetry requirements on the individual intra- and intersite terms will depend on the site and bond symmetry, respectively, and they will be addressed in the Appendix \ref{sec:app_symmM}. Crystal symmetries also determine the transformation rules for the coupling tensors of symmetry-equivalent sites and bonds. When inversion symmetry is a space-group element of the crystal, each site/bond will have an inversion-partner, each carrying an equal and opposite intrasite/intersite contribution, given the polar character of both $\bm{\mathcal{P}}^l$ and $\bm{\mathcal{P}}^{lj}$. The corresponding intrasite contributions will clearly cancel out exactly. The symmetric intersite contribution will also cancel out, given the cosinusoidal (even) dependence on the helical spin modulation, leaving only the antisymmetric term depending instead on $\sin(\bm q\cdot\bm r_j)$. Therefore, the macroscopic polarization induced by a noncollinear helimagnetic configuration in centrosymmetric crystals can be univoquely expressed as in Eq. (\ref{eq.P_Mtensor}) in the main text.

Given Eq.~\ref{eq:general_P}, we can generalize Eq.~\ref{eq.atomisticDMI} as 
\begin{equation}
   {\mathcal P}^{ij}_{\alpha\beta\gamma} = \frac{\partial {\mathcal J}^{ij}_{\beta\gamma}}{\partial \mathcal{E}_\alpha } , 
\end{equation}
where $\mathcal{J}^{ij}_{\alpha\beta}$ is the tensorial form of the exchange interaction, describing the energy of two interacting spins $\mathbf{S}^i$ and $\mathbf{S}^j$ as $E_{ij}=\mathcal{J}^{ij}_{\alpha\beta}S^i_\alpha S^j_\beta$~\cite{RevModPhys.95.035004}. The DMI is the anti-symmetric part of $\mathcal{J}^{ij}_{\alpha\beta}$. That only the anti-symmetric part of ${\mathcal P}^{ij}_{\alpha\beta\gamma}$ gives a macroscopic polarization in centrosymmetric crystals then corresponds to only the DMI coupling linearly to a uniform electric field. Correspondingly, in non-centrosymmetric crystals, one may expect a linear coupling also between other components of $\mathcal{J}^{ij}_{\alpha\beta}$ to uniform electric fields.

\section{Symmetry requirements for atomistic magnetoelectric coupling tensors.}\label{sec:app_symmM}

Both the intrasite and intersite contributions to spin-induced polarization are expressed by polar rank-three tensors. As such, each of them is fully determined by 27 coefficients, that can be largely reduced by considering the appropriate symmetry operations, i.e., site symmetries for intrasite and bond symmetries for intersite term, respectively.

The intrasite coupling tensor $\bm{\mathcal{P}}^i$ is a polar rank-three tensor symmetric with respect to the exchange of the second and third Cartesian indices. From the symmetry point of view, it transforms exactly as the piezoelectric tensor, under the point-group operations defined by the set of site symmetries. If inversion is a site-symmetry operation, the intrasite coupling tensor will be exactly zero. The site symmetry of Cr atoms in CrI$_3$ monolayer is $D_3$, lacking inversion symmetry, and the following coefficients are symmetry allowed: $\mathcal{P}^i_{222}=-\mathcal{P}^i_{211}=-\mathcal{P}^i_{121}$, $\mathcal{P}^i_{123}=-\mathcal{P}^i_{213}$, where $1,2,3$ denotes $x,y,z$ Cartesian components in the reference frame shown in Fig. \ref{fig.CrI3_struct}. However, the two Cr atoms within the unit cell are related by inversion symmetry, and their intrasite contributions mutually cancel out for any helical spin configuration.

Bond symmetries will affect the symmetric and antisymmetric parts of the intersite coupling tensor differently, $(\bm{\mathcal{P}}^{ij})^S$ and $\bm{\mathcal{M}}^{ij}$. The symmetry-constrained form of the antisymmetric part has been already discussed in Ref. [\onlinecite{kaplan_prb2011}] for the set of bond symmetries originally devised by Moriya to discuss the form of the antisymmetric Dzyaloshinskii-Moriya interaction\cite{moriya1960}. Here we generalize such rules to include also the symmetric intersite coupling tensor. We denote $A$ and $B$ the two spin sites forming the dimer, and we choose without loss of generality the $\hat{x}$ axis parallel to the $AB$ bond, while we denote $O$ the point bisecting the bond.
\paragraph{Inversion symmetry.} If an inversion center exists at the point $O$, then inversion will cause an overall change of sign of the full coupling tensor as well as an exchange of sites $A\leftrightarrow B$. Hence ${\mathcal{P}}^{AB}_{\alpha\beta\gamma}=-{\mathcal{P}}^{BA}_{\alpha\beta\gamma}=-{\mathcal{P}}^{AB}_{\alpha\gamma\beta}$, which means that only the antisymmetric part of the tensor can be nonzero, without imposing any further constraint, recently also pointed out in Ref.~\onlinecite{solovyev2025}. According to Moriya's rules, in this case the Dzyaloshinskii vector $\bm D$ is exactly zero.
\paragraph{Mirror plane orthogonal to the bond.} The mirror plane is taken to cross the bisecting point $O$ and it is perpendicular to the bond direction. In the chosen reference frame, it can be denoted as $m_{100}$. As for the inversion symmetry, the mirror operation also exchanges sites $A\leftrightarrow B$, implying that symmetric and antisymmetric tensor elements are mutually exclusive. For the symmetric part we adopt Voigt notation, such that $e_{ijk}=e_{il}$ where $l=j$ if $j=k$ and $l=9-(j+k)$ if $j\neq k$. In this case, one finds:
\begin{eqnarray}
(\bm{\mathcal{P}}^{AB})^S&=&\left(\begin{array}{cccccc}
0 		& 0		 & 0		    &0 		 &  \mathcal{P}_{15}& \mathcal{P}_{16}\\
\mathcal{P}_{21} &  \mathcal{P}_{22}& \mathcal{P}_{23}	& \mathcal{P}_{24}& 0		& 0		\\
\mathcal{P}_{31} &  \mathcal{P}_{32}& \mathcal{P}_{33}  & \mathcal{P}_{34}& 0		& 0	
\end{array}\right)\\
\bm{\mathcal{M}}^{AB}&=&\left(\begin{array}{ccc}
\mathcal{M}_{11} &0 &0\\
0 &\mathcal{M}_{22}&\mathcal{M}_{23}\\
0& \mathcal{M}_{32}&\mathcal{M}_{33}
\end{array}\right)
\end{eqnarray}
All diagonal elements of the antisymmetric part are allowed by symmetry, whereas only the off-diagonal components with both Cartesian indices within the reflection plane can be non-zero. The symmetry-allowed components of the symmetric tensor are either those where the first Cartesian index is orthogonal to the reflection plane and only one among the second and third indices lies within it, or those displaying the first index within the reflection plane and the second and third indices coincide or both lie within the plane. 
The mirror symmetry is not broken by an electric field applied orthogonally to the bond, and the electric-field induced DMI of Eq. (\ref{eq.atomisticDMI}) displays only two components $D_2(\bm{\mathcal{E}}) = -\mathcal{M}_{22}\mathcal{E}_2-\mathcal{M}_{32}\mathcal{E}_3$ and $D_3(\bm{\mathcal{E}}) = -\mathcal{M}_{23}\mathcal{E}_2-\mathcal{M}_{33}\mathcal{E}_3$, consistently with the second Moriya's rule stating that the Dzyaloshinskii vector is perpendicular to the bond.
\paragraph{Mirror plane containing the bond.} For the sake of clarity we consider a mirror operation $m_{010}$, i.e., orthogonal to the $\hat{y}$ direction; the constrained form for an equivalent $m_{001}$ operation can be trivially deduced. This symmetry operation does not exchange the spin positions, and in general symmetric and antisymmetric parts will not be mutually exclusive. We get:
\begin{eqnarray}
(\bm{\mathcal{P}}^{AB})^S&=&\left(\begin{array}{cccccc}
\mathcal{P}_{11} &  \mathcal{P}_{12}& \mathcal{P}_{13}	& 0 		  &  \mathcal{P}_{15}& 0	    \\
0 		& 0		 & 0		    & \mathcal{P}_{24} & 0		& \mathcal{P}_{26}\\
\mathcal{P}_{31} &  \mathcal{P}_{32}& \mathcal{P}_{33}  & 0 		 & \mathcal{P}_{35}		& 0	
\end{array}\right)\\
\bm{\mathcal{M}}^{AB}&=&\left(\begin{array}{ccc}
0 & \mathcal{M}_{12} &0\\
\mathcal{M}_{21} &0&\mathcal{M}_{23}\\
0& \mathcal{M}_{32}& 0
\end{array}\right)
\end{eqnarray}
The antisymmetric part will have no diagonal elements, while off-diagonal elements are nonzero if only one of the Cartesian indices is perpendicular to the mirror. For diagonal elements of the symmetric part, $(\mathcal{P}^{AB})^S_{\alpha\beta\beta}$, all those elements where the first index $\alpha$ is orthogonal to the reflection plane are zero; off-diagonal elements are zero if the first index $\alpha$ is within the reflection plane and both the other indices $\beta,\gamma$ are within the plane, or viceversa (i.e., if the first index is orthogonal to the reflection plane, only the element with both the second and third indices within the reflection plane is nonzero). This case is compatible with an electric field applied within the reflection plane. For the $m_{010}$ reflection, we get a single component for the induced DMI, $D_2(\bm{\mathcal{E}})=-\mathcal{M}_{12}\mathcal{E}_1-\mathcal{M}_{32}\mathcal{E}_3$, that is orthogonal to the reflection plane in agreement with the third Moriya's rule. 
\paragraph{Two-fold rotation axis orthogonal to the bond.} Without loss of generality, we consider a two-fold rotation axis passing through $O$ and parallel to the $\hat{y}$ axis orthogonal to the bond, $2_{010}$. This operation swaps the site positions, implying mutually exclusive symmetric and antisymmetric parts. We get:
\begin{eqnarray}
(\bm{\mathcal{P}}^{AB})^S&=&\left(\begin{array}{cccccc}
0 		& 0		 & 0		&  \mathcal{P}_{14} 		  & 0&  \mathcal{P}_{16}	    \\
\mathcal{P}_{21} &  \mathcal{P}_{22}& \mathcal{P}_{23}	  & 0 &  \mathcal{P}_{25}   &  0 \\
0 		& 0		 & 0	  &  \mathcal{P}_{34}		 & 0		&  \mathcal{P}_{36}	
\end{array}\right)\\
\bm{\mathcal{M}}^{AB}&=&\left(\begin{array}{ccc}
0 & \mathcal{M}_{12} &0\\
\mathcal{M}_{21} &0&\mathcal{M}_{23}\\
0& \mathcal{M}_{32}& 0
\end{array}\right)
\end{eqnarray}
Notice that the antisymmetric part is exactly the same as for case $c$. This is not surprising, as the mirror operations of case $c$ can be obtained as the twofold rotations considered here followed by inversion, that does not enforce any additional condition on the $\bm{\mathcal{M}}$ matrix elements. Summarizing, the latter will be nonzero if only one Cartesian index is parallel to the rotation axis. The diagonal elements in the second pair of indices of the symmetric part are nonzero if the first Cartesian index is along the rotation axis, while the only nonzero off-diagonal elements have only one Cartesian coordinate along it. An electric field applied along the $2_{010}$ axis would induce a DMI with components $D_1(\bm{\mathcal{E}})=-\mathcal{M}_{21}\mathcal{E}_2$ and $D_3(\bm{\mathcal{E}})=-\mathcal{M}_{23}\mathcal{E}_2$, i.e., orthogonal to the twofold rotation axis as required by the fourth Moriya's rule.
\paragraph{n-fold rotation axis along the bond.} We consider first the case of two-fold rotation, $2_{100}$, and then generalize to any $n>2$. This operation does not swap the atomic positions. The symmetry-allowed tensor elements are:
\begin{eqnarray}
(\bm{\mathcal{P}}^{AB})^S&=&\left(\begin{array}{cccccc}
\mathcal{P}_{21} &  \mathcal{P}_{22}& \mathcal{P}_{23}	    &\mathcal{P}_{14} 		 & 0 & 0	\\
0 		& 0		 & 0	 		& 0& \mathcal{P}_{25}		& \mathcal{P}_{26}		\\
0 		& 0		 & 0	  		&0& \mathcal{P}_{35}		& \mathcal{P}_{36}	
\end{array}\right)\\
\bm{\mathcal{M}}^{AB}&=&\left(\begin{array}{ccc}
\mathcal{M}_{11} &0 &0\\
0 &\mathcal{M}_{22}&\mathcal{M}_{23}\\
0& \mathcal{M}_{32}&\mathcal{M}_{33}
\end{array}\right)
\end{eqnarray}
For $n>2$ the higher symmetry further reduces the number of independent coefficients, being $\mathcal{P}_{26}=\mathcal{P}_{35}$ and $\mathcal{P}_{25}=-\mathcal{P}_{36}$, as well as $\mathcal{M}_{23}=-\mathcal{M}_{32}$. The DMI induced by an electric field parallel to the bond is described by a single component $D_1=-\mathcal{M}_{11}\mathcal{E}_1$, i.e., the Dzyaloshinskii vector is parallel to the bond direction consistently with the fifth Moriya's rule.

We remark that all above rules apply to cubic systems, recovering the KNB result for the spin-current contribution\cite{PhysRevLett.107.157202}. Note that this is not applicable in any 2D magnet. In CrI$_3$ the first nearest-neighbor bonds relate atomic sites belonging to two sublattices related by inversion symmetry, and additionally display a mirror plane orthogonal to the bonding vector. The symmetric part of the coupling tensor is exactly zero, while the antisymmetric part obeys the second rule $b)$ above. Exactly the same symmetry requirements apply to the third nearest neighbor. For both cases, the $\bm{\mathcal{M}}$ matrix for bonding vectors parallel to the $\hat{y}$ axis takes the form of Eq. (\ref{eq:M1_M3}), the other symmetry-equivalent bonds being related by the three-fold rotation axis orthogonal to the monolayer. The second nearest neighbor bonds instead relate atomic sites belonging to the same sublattice, thus lacking inversion symmetry and displaying a two-fold rotation axis orthogonal to the bonding vector. According to fourth rule $d)$, the $\bm{\mathcal{M}}$ matrix for bonding vectors parallel to the $\hat{x}$ axis takes the form of Eq. (\ref{eq:M2}), and the matrix form of symmetry-equivalent bonds can be obtained by applying the threefold rotations $3^{\pm}_{001}$ and an in-plane twofold rotation. Since second nearest-neighbor bonds are acentric, the symmetric part of the coupling tensor will in general be nonzero. Calculated values for CrI$_3$ are one order of magnitude smaller than the dominant antisymmetric components, the largest being $\mathcal{P}_{14} = 14\cdot 10^{-5}~\mathrm{e\AA}$ and  $\mathcal{P}_{25} = 28\cdot 10^{-5}~\mathrm{e\AA}$, while the remaining are less than $1\cdot 10^{-5}~\mathrm{e\AA}$ and too small to be determined with sufficient numerical accuracy. Given the global inversion symmetry of CrI$_3$, $(\bm{\mathcal P}^{2})^S$ will not contribute to the macroscopic polarization induced by magnetic spirals, as discussed in Appendix \ref{sec:app_macroP}; for localized magnetic defects, the symmetric part of the coupling tensor can instead contribute to the spatial modulation of the electric polarization, providing corrections that are of higher order in the atomic spacing, that we neglect for the spatially extended magnetic modulations considered in this work.

\section{Symmetric form of $f_{\alpha\beta,\delta\gamma}$}\label{app.contmodel}

The tensor $f_{\alpha\beta,\gamma\delta}$ appearing in Eq.~\ref{eq.contmod} is a fourth rank tensor with up to 81 elements in the 3D case, or 54 in 2D. These are reduced by symmetries as $f_{\alpha\beta,\gamma\delta}$ should be invariant under the crystallographic point group symmetries. 
Similarly to the atomistic case treated above, $f_{\alpha\beta,\gamma\delta}$ can be decomposed in symmetric and anti-symmetric parts 
$f_{\alpha\beta,\gamma\delta} = \frac{1}{2}\left(f_{\alpha\beta,\gamma\delta} + f_{\alpha\beta,\delta\gamma} \right) + \frac{1}{2}\left( f_{\alpha\beta,\gamma\delta} - f_{\alpha\beta,\delta\gamma} \right) \equiv f_{\alpha\beta,\gamma\delta}^\mathrm{sym} + \tilde{f}_{\alpha\beta,\gamma\delta}$
As pointed out by Mostovoy~\cite{PhysRevLett.96.067601}, only $\tilde{f}_{\alpha\beta,\gamma\delta}$ contributes to the macroscopic polarization, or net electric dipole moment. This is easily seen by doing an integration by parts of Eq.~\ref{eq.contmod}. The anti-symmetric tensor $\tilde{f}_{\alpha\beta,\gamma\delta}$ has up to 27 elements in 3D and 18 in 2D.  The symmetric part of the tensor is symmetry-wise equivalent to the flexoelectric tensor.

The rank four tensor $\tilde{f}_{\alpha\beta,\gamma\delta}$, that is anti-symmetric in the last two indices, has a single independent, non-zero component $\tilde{f}_{12,12}$ for the three highest symmetry crystallographic point groups $O_h$, $T_d$ and $O$~\cite{Gallego:lk5043}. These are the points groups for which the model studied by Mostovoy~\cite{PhysRevLett.96.067601} is valid. While this is consistent with the KNB model, in the atomistic case local bond symmetries should be considered. Additionally, there are 13 crystallographic point groups, $T_h$, $T$, $D_{6h}$, $D_{3h}$, $C_{6v}$, $D_6$, $D_{4h}$, $D_{2d}$, $C_{4v}$, $C_4$, $D_{2v}$, $C_{2v}$, $D_{2}$, for which the reduced symmetry only breaks the equivalence between some elements in $\tilde{f}_{\alpha\beta,\gamma\delta}$, without allowing for new non-zero ones. The 16 remaining points groups are those for which spin spirals are expected to induce additional magnetoelectric polarization components than those expected from the high symmetry models considered by Mostovoy~\cite{PhysRevLett.96.067601} and KNB~\cite{KNB}.

\begin{table}[h]
\begin{tabular}{ l }
\hline  
$f_{11,33}  = f_{22,33} $ 	 \\  
$f_{22,22}  = f_{11,11} = f_{12,21} + f_{21,21} + f_{22,11} $   \\ 
$f_{12,21}  = f_{21,12} $  \\ 
$f_{12,12}  = f_{21,21} $  \\ 
$f_{11,22}  = f_{22,11} $   \\ 
$f_{31,22}  = f_{32,12} = f_{32,21} = -f_{31,11}  $    \\ 
$f_{11,31}  = - f_{21,32} = - f_{12,32} = -f_{22,31} $  \\ 
$f_{11,13}  = -f_{21,23} = -f_{12,23} = -f_{22,13} $  \\ 
$ f_{31,31}  = f_{32,32}  $  \\ 
$ f_{31,13}  = f_{32,23}  $ \\ 
\hline
\end{tabular}
\caption{The 9 independent, non-zero components of $f_{\alpha\beta,\gamma\delta}$ and their relations.}
\label{table.fijab}
\end{table}
The $D_{3d}$ symmetry of the CrI$_3$ monolayer sets 28 components to zero and leaves 9 independent, non-zero components, listed in Table~\ref{table.fijab}. 
Anti-symmetrizing the tensor, leaves only three independent, non-zero elements
\begin{align}
    \tilde{f}_\mathrm{ip} \equiv \tilde{f}_{11,31} = & \tilde{f}_{21,23} = \tilde{f}_{12,23} = \tilde{f}_{22,13} \\
    \tilde{f}_{z} \equiv \tilde{f}_{31,31} = & \tilde{f}_{32,32}    \\ 
    \tilde{f}_{12,12} = & \tilde{f}_{21,21}. 
\end{align}

\section{Skyrmion Solution with Electric Field/DMI}\label{app.Skyrm_sol}
The total magnetic energy functional, including electric field-induced DMI, is obtained by adding $\mathcal{D} = -\mathbf{\mathcal{E}}\cdot\mathbf{P}(\mathbf{r})$ to the energy density in Eq.~\ref{eq.E_skyrm_nofield} (for $\nu=\pm1$)
\begin{widetext}
\begin{align}
E  & = 2\pi A \int_0^\infty r \left(    [ \Theta^\prime (r) ]^2 +\left( \frac{1}{r^2} + \frac{1}{\xi^2} \right)\sin^2\Theta - \left| \frac{\mathcal{E}_i}{\eta_i} \right|  \frac{1}{\xi}\left[ \Theta^\prime(r) + \frac{\sin2\Theta}{2r}\right] \right)  \dd r  \label{eq.E_Efield}  \\ 
   & = \underbrace{ 2\pi A \int_0^\infty \rho   \left( [ \Theta^\prime (\rho) ]^2 +\frac{1}{\rho^2}\sin^2\Theta \right) \dd \rho }_{a} + \underbrace{ 2\pi  K \int_0^\infty \rho   \sin^2\Theta  \dd \rho }_{k} w^2  -  |\mathcal{E}_i | \underbrace{ 2\pi  \left| \tilde{f}_{i} \right|   \int_0^\infty \rho  \left( \Theta^\prime(\rho) + \frac{{\sin2\Theta}}{2\rho}\right) \dd \rho }_{d_i} w 
   \label{eq.E_Efield_rho} 
\end{align}
\end{widetext}
where $\tilde{f}_i$ is either $\tilde{f}_\mathrm{ip}$ or $\tilde{f}_{z}$, depending on whether the electric field is oriented in or out-of-plane. Here we introduce the direction dependent electric field unit $\eta_i = \frac{A}{|f_i| \xi} = \frac{\sqrt{AK}}{|f_i|}$. In Eq.~\ref{eq.E_Efield_rho} we made the variable substitution $\rho = r/w$ to illustrate the meaning of $a$, $d$ and $k$. The helicity $\gamma$ and polarity $\mathcal{P}=\pm1$ are assumed to adapt themselves to maximize the gain in electric dipole energy, i.e. $(-\cos\gamma,\sin\gamma,0)$ will be (anti-)parallel to the field if it is in-plane, or $\gamma$ will be either $0$ or $\pi$ if it is out-of-plane. There is no unique combination of values that gives the ground state solution, since changing simultaneously both $\mathcal{P}$ and $\gamma$ gives different degenerate states. The Euler-Lagrange equation that minimizes Eq.~\ref{eq.E_Efield} is 
\begin{equation}\label{eq.EL}
\xi^2 \Theta^{\prime\prime}(r) + \frac{\xi^2}{r}\Theta^{\prime}(r) - \frac{1}{2}\sin2\Theta \left( 1 + \frac{\xi^2}{r^2}\right) + \frac{\mathcal{E}_i}{\eta_i} \frac{\xi}{r} \sin^2\Theta = 0.
\end{equation}
Since this equation depends only on $\xi$ and the value of the electric field relative to $\eta_i$, Skyrmion solutions for out-of-plane electric fields and anti-Skyrmion solutions for in-plane fields will have identical energy and radial profile $\Theta(r)$ at the same DMI values of $D_i=\left| f_{i} \right| \mathcal{E}_i$. By numerically solving Eq.~\ref{eq.EL} we find the radial profile $\Theta(r)$ and the corresponding energies as functions of electric field, as plotted in Fig.~\ref{fig.Skyrm_sols}. Note that we only consider cases with either in-plane or out-of-plane fields, not intermediate directions. 

Given the ansatz $\Theta({r}) = 4\arctan(\mathrm{e}^{-r})$, we can evaluate the integrals in Eq.~\ref{eq.E_Efield_rho} to
\begin{align}
a & = 2\pi A \left( 4 \ln 2 + I \right) \\ 
k & = \frac{4\pi}{3} K \left( 1 + \ln 4\right) \\ 
d_i & = 2\pi f_i \left( \frac{2}{3} - 4 C \right), 
\end{align}
where 
\begin{align}
    I & = \int_0^\infty \frac{\sin^2 \Theta}{r} \mathrm{d}r \approx 1.53971 \\ 
    C & = \sum_{k=0}^{\infty} \frac{ (-1)^k }{(2k + 1)^2} \approx 0.91597.
\end{align}

\bibliography{literature}%

\end{document}